\documentclass[useAMS,usenatbib]{mn2e}

\usepackage[final]{pdfpages}
\usepackage{longtable}
\usepackage{graphicx}
\usepackage{epsfig}
\usepackage{lscape}
\usepackage{tablefootnote}
\usepackage{xcolor}
\usepackage[breaklinks,colorlinks,citecolor=blue]{hyperref}
\usepackage{breakurl}
\usepackage{caption}
\usepackage{subfig}
\DeclareCaptionLabelFormat{continued}{Figure #1 #2. (continued)} 
\newcommand{\ha}{H$\alpha$} 
\newcommand{\hi}{H{\sc i}} 
\newcommand{\hii}{H{\sc ii}} 
\newcommand{\paper}{Paper {\sc ii}}
\newcommand{\pap}{Paper {\sc i}}

\def\arcm{\ifmmode {^{\scriptstyle\prime}}
          \else $^{\scriptstyle\prime}$\fi}
\def\farcm{\ifmmode $\setbox0=\hbox{$^{\prime}$}\rlap{\hskip.11\wd0 .}$^{\prime}
          \else \setbox0=\hbox{$^{\prime}$}\rlap{\hskip.11\wd0 .}$^{\prime}$\fi}
\def\pdeg{\ifmmode $\setbox0=\hbox{$^{\circ}$}\rlap{\hskip.11\wd0 .}$^{\circ}
          \else \setbox0=\hbox{$^{\circ}$}\rlap{\hskip.11\wd0 .}$^{\circ}$\fi}
\def\vhel{\ifmmode{V_{{\rm HEL}}}\else{$V_{{\rm HEL}}$}\fi}
\def\vsys{\ifmmode{V_{\rm sys}}\else{$V_{\rm sys}$}\fi}
\def\kms{\ifmmode{\,{\rm km\,s}^{-1}}\else{\,km\,s$^{-1}$}\fi}
\def\vlsr{\ifmmode{v_{\rm lsr}}\else{$v_{\rm lsr}$}\fi}
\def\ltsim{\ifmmode\stackrel{<}{_{\sim}}\else$\stackrel{<}{_{\sim}}$\fi}
\def\gtsim{\ifmmode\stackrel{>}{_{\sim}}\else$\stackrel{>}{_{\sim}}$\fi}

\def\reff@jnl#1{{\rm#1\/}}
\def\aj{\reff@jnl{AJ}}                  
\def\araa{\reff@jnl{ARA\&A}}            
\def\apj{\reff@jnl{ApJ}}                
\def\apjl{\reff@jnl{ApJ}}               
\def\apjs{\reff@jnl{ApJS}}              
\def\ao{\reff@jnl{Appl.Optics}}         
\def\apss{\reff@jnl{Ap\&SS}}            
\def\aap{\reff@jnl{A\&A}}               
\def\aapr{\reff@jnl{A\&A~Rev.}}         
\def\aaps{\reff@jnl{A\&AS}}             
\def\azh{\reff@jnl{AZh}}                        
\def\baas{\reff@jnl{BAAS}}              
\def\jrasc{\reff@jnl{JRASC}}            
\def\memras{\reff@jnl{MmRAS}}           
\def\mnras{\reff@jnl{MNRAS}}            
\def\pra{\reff@jnl{Phys.Rev.A}}         
\def\prb{\reff@jnl{Phys.Rev.B}}         
\def\prc{\reff@jnl{Phys.Rev.C}}         
\def\prd{\reff@jnl{Phys.Rev.D}}         
\def\prl{\reff@jnl{Phys.Rev.Lett}}      
\def\pasp{\reff@jnl{PASP}}              
\def\pasj{\reff@jnl{PASJ}}              
\def\qjras{\reff@jnl{QJRAS}}            
\def\skytel{\reff@jnl{S\&T}}            
\def\solphys{\reff@jnl{Solar~Phys.}}    
\def\sovast{\reff@jnl{Soviet~Ast.}}     
 \def\ssr{\reff@jnl{Space~Sci.Rev.}}     
\def\zap{\reff@jnl{ZAp}}                        
\def\nat{\reff@jnl{Nature}}             

\def\LaTeX{L\kern-.36em\raise.3ex\hbox{a}\kern-.15em
    T\kern-.1667em\lower.7ex\hbox{E}\kern-.125emX}

\def\deg{^\circ}

\begin{document}

\title[The HIPASS RRL survey]{The HIPASS survey of the Galactic plane in Radio Recombination Lines}
\author[M.I.R. Alves et al.]{Marta
  I. R. Alves,$\!^{1,2}$\thanks{E-mail:marta.alves@irap.omp.eu} Mark
  Calabretta,$\!^{3}$ Rodney D. Davies,$\!^{2}$ Clive
  Dickinson,$\!^{2}$\newauthor Lister Staveley-Smith,$\!^{4}$ Richard J.
  Davis,$\!^{2}$ Tianyue Chen$^{2}$ and Adam Barr$^{2}$\\
$^1$Institut d'Astrophysique Spatiale, CNRS (8617) Universit\'{e} Paris-Sud 11, B\^{a}timent 121, Orsay, France\\
$^2$Jodrell Bank Centre for Astrophysics, Alan Turing Building, School of Physics and Astronomy, \\
The University of Manchester, Oxford Road, Manchester, M13 9PL, UK \\
$^3$CSIRO Astronomy and Space Science, PO Box 76, Epping, NSW 1710, Australia\\
$^4$International Centre for Radio Astronomy Research, M468, University of Western Australia,\\
35 Stirling Hwy, Crawley, WA 6009, Australia\\
}

\date{Accepted 2015 April 2. Received 2015 March 27; in original form 2014 November 17}
       
\pagerange{\pageref{firstpage}--\pageref{lastpage}} 
\pubyear{}

\maketitle
\label{firstpage}


\begin{abstract}

We present a Radio Recombination Line (RRL) survey of the Galactic
Plane from the \hi~Parkes All-sky Survey and associated Zone
of Avoidance survey, which mapped the region $l=196\degr$ -- $0\degr$ -- $52\degr$ and $|b| \leq
5\degr$ at 1.4\,GHz and 14.4\,arcmin resolution. We combine three RRLs,
H168$\alpha$, H167$\alpha$, and H166$\alpha$ to derive fully
sampled maps of the diffuse ionized emission along the inner
Galactic plane. The velocity information, at a resolution of 20\kms,
allows us to study the spatial distribution of the ionized gas and
compare it with that of the molecular gas, as traced by CO. The
longitude-velocity diagram shows that the RRL emission is mostly
associated with CO gas from the molecular ring and is concentrated
within the inner 30\degr\ of longitude. A map of the free-free emission in this region of the Galaxy is
derived from the line-integrated RRL
emission, assuming an electron temperature gradient with Galactocentric radius of
$496\pm100$\,K\,kpc$^{-1}$. Based on the
thermal continuum map we extracted a catalogue of 317 compact (\ltsim
15\,arcmin) sources, with
flux densities, sizes and velocities. We report the first RRL
observations of the southern ionized lobe in the Galactic centre. The
line profiles and velocities suggest that this degree-scale structure
is in rotation. We also present new
evidence of diffuse ionized gas in the 3-kpc arm. Helium and carbon RRLs
are detected in this survey. The He line is mostly observed
towards \hii~regions, whereas the C line is also detected further away
from the source of ionization. These data represent the first observations of
diffuse C RRLs in the Galactic plane at a frequency of 1.4\,GHz. 

\end{abstract}

\setcounter{figure}{0}

\section{INTRODUCTION}
\label{sec:introduction}

Radio Recombination Lines (RRLs) have been widely used in astrophysics, especially as
probes of the physical conditions of the line emitting plasma and
\hii~regions. From the observation of a single transition, the
electron temperature of the gas can be determined by comparing the
energy radiated by the line to that of the underlying thermal
free-free continuum. This remains one of the most accurate methods to
determine the electron temperature. The central velocity of the line
and its width give information about the systematic and turbulent
motions of the gas. 

In this paper we present the RRL data from the
\hi~Parkes All-Sky Survey (HIPASS, \citealt{Staveley-Smith:1996}) and associated Zone of
Avoidance Survey (ZOA, \citealt{Staveley-Smith:1998}), which cover the Galactic
plane accessible from Parkes, $l=196\degr$ -- $0\degr$ --
  $52\degr$, and $|b| \leq
5\degr$. The
first results from the analysis of the three H$\alpha$ RRLs,
H168$\alpha$, H167$\alpha$, and H166$\alpha$, for a smaller
section of the Galactic plane are given in
\citet{Alves:2010,Alves:2012} (hereafter \pap~and
\paper\ respectively). \citet{Calabretta:2014} used the
full HIPASS and ZOA datasets to produce a map of the 1.4\,GHz continuum emission in
the southern sky.

RRLs arising from low density gas are relatively weak, with
  temperatures of tens to a
hundred mK at frequencies near 1\,GHz, thus about 1 per cent of the
thermal continuum brightness in the Galactic plane (see Section
\ref{sec:fftemp}).
So far studies of the diffuse ionized
gas have been performed by pointed observations towards positions in the
inner plane free from strong \hii~regions
(e.g., \citealt{Gottesman:1970}, \citealt{Jackson:1971}, \citealt{Heiles:1996},
\citealt{Baddi:2012}). Here we present the first contiguous survey of RRL
emission in the plane of the Galaxy, suitable for a comprehensive
study of the diffuse ionized medium. We note that similar surveys, at
the same frequency and at higher angular and spectral resolution, are
underway with the Arecibo telescope (SIGGMA, \citealt{Liu:2013}) and
the Very Large Array (THOR, \citealt{THOR:2013}). The former will
cover part of the first quadrant and a region of the anti-centre,
$l=30\degr$ -- $75\degr$ and $l=175\degr$ -- $207\degr$ respectively,
within $2\degr$ of the Galactic plane, whereas the latter will map the
region $l=15\degr$ to $67\degr$, $|b|\leq 1\degr$.

Our previous RRL study of the region $l=20\degr$ -- $44\degr$ and $|b| \leq
4\degr$ revealed a width of the emission at velocities
corresponding to the Sagittarius and Scutum spiral arms of $0\pdeg85
\pm 0\pdeg03$ and $0\pdeg78 \pm 0\pdeg01$ respectively (\paper). This
is equivalent to a $z$-thickness of $96 \pm 4$\,pc and thus slightly
broader than 74\,pc, the width of the distribution of OB stars, the
progenitors of \hii~regions \citep{Bronfman:2000}. This implies that
the ionizing radiation from the OB stars is not fully absorbed in the
dense molecular clouds in which they are born, but escapes into the
surrounding lower density medium, which it ionizes.

The inner Galaxy has been mapped and studied over a wide range of
frequencies, from radio to infra-red (e.g.,
\citealt{Haynes:1978, Haslam:1982, Reich:1982, Reich:1990a,
  Miville-Deschenes:2005, Stil:2006, Jarosik:2011,
  Sun:2011}). \citet{Paladini:2005} and \citet{Sun:2011} have used
some of these data to separate the thermal and non-thermal  
emission at radio frequencies by exploiting the difference in their brightness
temperature spectral indices. Even if the free-free spectrum is well
known, uncertainties in the synchrotron spectral behaviour and degree
of polarization, as well as in the zero levels of the 
different surveys, limit this method. However, a more direct approach is
possible thanks to the RRL data. These provide a measure of the
free-free emission that, when compared with the total continuum at the
same frequency, enables the separation of the synchrotron component (\paper).

One obvious advantage of RRLs over other thermal
emission tracers is that they are not attenuated by interstellar dust
along the line of sight and can thus be interpreted directly. This
provides an unambiguous estimate of the Emission Measure (${\rm EM}
= \int n_{e}^2 d\ell$), which is valuable for the study of the electron
density distribution.
The RRL emission from ionized hydrogen gas can be expressed
in terms of the integral over spectral line temperature, $T_{\rm L}$, as
\begin{equation}
\int T_{\rm L}  d \nu  = 1.92 \times 10^{3}  T_{\rm e}^{-1.5}  {\rm EM} 
\label{eq:1}
\end{equation}
where $\nu$ is the frequency (kHz), $T_{\rm e}$ is the electron
temperature in K and the
emission measure EM is in cm$^{-6}$\,pc \citep{Rohlfs:2000}.
The corresponding continuum
emission brightness temperature is
\begin{equation}
T_{\rm b} = 8.235 \times 10^{-2} a(T_{\rm e},\nu)  T_{\rm e}^{-0.35} \nu^{-2.1}_{\rm GHz}
 (1+0.08)  {\rm EM}
\label{eq:2}
\end{equation}
where $a(T_{\rm e},\nu)$ is a
slowly varying function of the electron temperature and frequency
\citep{Mezger:1967, D3:2003,Draine:2011} and the $(1+0.08)$ factor represents the
additional contribution to $T_{\rm b}$ from helium.
Thus, the underlying free-free emission can be obtained from the
 integrated RRL as follows
\begin{equation}
T_{\rm b} = \frac{1}{6.985 \times 10^{3}} a(T_{\rm e},\nu) (1+0.08)
T_{\rm e}^{1.15} \nu_{\rm GHz}^{-1.1} \int{T_{\rm L} dV}
\label{eq:ff}
\end{equation}
where $V$ is in \kms. This equation is
valid for $\alpha$ transitions (i.e., one-level transitions), in Local Thermodynamic
Equilibrium \citep{Rohlfs:2000}. 

The present data allow the determination of a fully-sampled map of the free-free emission
in the inner Galaxy, thus providing an even better sampling of the
spiral arms. Such a map is valuable not only for separating thermal
and non-thermal emission, allowing synchrotron objects to be
studied at 1.4\,GHz, but also for star formation studies
since it includes both the individual and bright \hii~regions and the
extended ionized emission surrounding them. 
Recently these RRL data
have been used as part of an inversion method that separates dust
emission in the \textit{Herschel} bands associated with atomic, molecular and ionized gas in a
$2\degr \times 2\degr$ region centred at $(l,b)=(30\degr,0\degr)$ \citep{Traficante:2014}. These data are unique in that they provide
the velocity information of the ionized component necessary for this
work and until now unavailable in the Galactic
plane. \citet{Tibbs:2012} used the RRL data towards the Anomalous
Microwave Emission (AME) object RCW175 to confirm its thermal radio
component and also to compare the velocity of the ionized gas with
that of other gas tracers. 
Furthermore, the separation of the several foreground emission components
from Cosmic Microwave Background observations, such as those
delivered by the \textit{Planck} satellite, can be improved significantly
using the results reported here. An estimate of the free-free emission in the
Galactic plane is a major input to disentangle the other components at
radio frequencies. \citet{PIP79} used the free-free map of \paper\ to
isolate the AME in the Galactic plane region $l=20\degr$ -- $40\degr$and
$|b| \leq 4\degr$. The same data have also been used by \citet{PIP96} to remove the ionized emission from the millimetre channels of \textit{Planck} and thus to study and characterise the emission of interstellar dust.

This paper is organised as follows. Section \ref{sec:rrldata}
describes the RRL survey, data reduction and calibration. In Section
\ref{sec:fftemp} we present the RRL maps and study the
spatial distribution of the ionized gas, namely how it compares with
that of the molecular gas. We also show evidence of carbon RRL
emission from the diffuse medium in the Galactic plane.
Section \ref{sec:hiicat} presents a list of \hii~regions extracted from
the free-free map, along with their flux densities at 1.4\,GHz,
angular sizes and line velocities. Section~\ref{sec:gcl} focuses on the
Galactic centre region, where we study the structure of the ionized
degree-scale lobe and present observations of diffuse emission from the
3-kpc arm. Finally, we conclude with a summary of the present
  work in Section~\ref{sec:conc}.

\section{THE RRL SURVEY AND DATA}
\label{sec:rrldata}

The HIPASS and ZOA surveys are described in
\citet{Staveley-Smith:1996, Staveley-Smith:1998}. These searched for
\hi-emitting galaxies in the velocity range $-1280$ to 12700\kms~using
the Parkes 21-cm multibeam receiver. Within the 64\,MHz bandwidth there
are three RRLs: H168$\alpha$, H167$\alpha$ and H166$\alpha$ at
1374.601, 1399.368 and 1424.734\,MHz, respectively, which are combined
and used to recover the thermal emission. The HIPASS survey covers the
sky south of $\delta=+25\pdeg5$. The multibeam receiver has 13 beams,
each with two linear polarizations (Stokes $I$ only), set on a hexagonal grid with a
mean observing beam width of 14.4\,arcmin full width at half maximum
(FWHM). The survey is divided into 15 zones and the scans are taken at
constant right ascension, in declination strips of
8\degr~separated by 7\,arcmin. Hence, each of the 13 beams mapped the sky
at slightly below the Nyquist rate. The ZOA deep survey scanned at
constant Galactic latitude and each $8\pdeg5$ longitude scan is separated by
1.4\,arcmin. It covers most of the Galactic plane accessible from Parkes,
$l=196\degr$ -- $0\degr$ -- $52\degr$ and $|b| \leq 5\degr$, and is
divided into 27 zones. The
latitude coverage was extended to $|b| \leq 10\degr$ for $l=332\degr$ -- $0\degr$ -- $36\degr$ and
$|b| \leq 15\degr$ for $l=348\degr$ -- $0\degr$ -- $20\degr$. The
footprint of the receiver on the sky has a width of $1\pdeg7$, thus
each scan maps an area slightly greater than $8\pdeg5 \times 1\pdeg7$. The multibeam
correlator cycle is 5\,s and the scan rate is 1\degr\,min$^{-1}$, therefore
each scan has 100 integrations. The HIPASS/ZOA frequency coverage is
from 1362.5 to 1426.5\,MHz centred on 1394.5\,MHz and is divided into
1024 channels spaced by 62.5\,kHz, or 13.4\kms~for the H167$\alpha$~line. The
total integration time is 450\,s per beam for the HIPASS survey and
2100\,s per beam for the five times deeper ZOA survey. This results in
a typical rms (root mean square) noise of 13\,mJy\,beam$^{-1}$ and
6\,mJy\,beam$^{-1}$ for the HIPASS and ZOA surveys, respectively.

The results presented in this paper use a combination of the HIPASS
and ZOA datasets for the Galactic plane region $l=196\degr$ -- $0\degr$
-- $52\degr$ and $|b| \leq 5\degr$. Table~\ref{table1} lists the
relevant observing parameters. The data analysis techniques and
properties of the final maps are described below.

\begin{table}
\centering
\caption{Summary of the HIPASS/ZOA RRL survey parameters.\label{table1}}
\begin{tabular}{lc}
\hline
\hline
Beams & 13 \\
           & 1 central, 6 inner ring, 6 outer ring\\
Beam FWHM & 14.0, 14.1, 14.5 arcmin \\
Beam ellipticity & 0.00, 0.03, 0.06 \\
Gridded beam FWHM & 14.4 arcmin \\
Polarizations & 2 orthogonal linear \\
                      & (Stokes $I$ only) \\
Longitude coverage & $l=196\degr$ -- $0\degr$ -- $52\degr$ \\
Latitude coverage &  $|b| \leq 5\degr$\\
RRLs included: & \\
H168$\alpha$ & 1374.601\,MHz \\
H167$\alpha$ & 1399.368\,MHz  \\
H166$\alpha$ & 1424.734\,MHz \\
Scan rate & 1\degr\,min$^{-1}$ \\
Integration time & 5\,s \\
Channel separation$^{\rm a}$ & 13.4\kms\ \\
Velocity resolution & 20\kms\ \\
rms noise$^{\rm b}$ & 4.5\,mK (6.4\,mJy\,beam$^{-1}$) \\
$T_{\rm b}/S$ & $0.70\pm0.07$\,K\,/(Jy\,beam$^{-1}$) \\
\hline
\end{tabular}
$^{\rm a}$ Equivalent to 62.5\,kHz at the frequency of the
H167$\alpha$ line. \\
$^{\rm b}$ Typical rms noise in the stacked spectra, at a
  velocity resolution of 20\kms, along the
  Galactic plane.
\end{table}

\subsection{Data analysis}
\label{sec:datanalaysis}

In this section we describe the main aspects of the data
analysis, which have been presented in \paper. There are two
additional steps, related to the continuum
ripple removal and gridding of the data, which are detailed in
\citet{Calabretta:2014}, since the reduction of the HIPASS/ZOA
spectral data follows closely that of the continuum. 

The spectra generated by the multibeam correlator are dominated by the system
bandpass, which is corrected using the
\textit{tsysmin} algorithm (\paper) in the
\textit{livedata}\footnote{www.atnf.csiro.au/computing/software/livedata.html}
software \citep{Barnes:2001}. \textit{tsysmin}, which was developed to recover extended
emission in the Galactic plane, locates the $n$ ($=10$ in this work) integrations in
each scan for which the median value of the system temperature is a minimum and, for each
spectral channel, takes the median value over the same integrations as
the bandpass correction. Other effects in the spectra such as spectral
ringing due to strong \hi~Galactic emission, baseline ripple, and non-linearities in the receiver and amplifier system are not
eliminated in this process. Spectral ringing is reduced by smoothing the
spectra using the Tukey filter \citep{Tukey:1967}. Baseline ripple is
caused by single or multiple reflections of radiation from strong
continuum sources between the dish and the receiver. Even though its
pattern is not purely sinusoidal, due to reflections from the feed
legs as well, it can be well reproduced and removed from the spectra. In
addition to the ripple, a rise in the bandpass response towards
low-frequencies is observed in the spectra of strong continuum
sources, proportional to their strength. These two effects are
corrected via a scaled template method: a normalized continuum
baseline signature is determined for each beam and polarization using
a large subset of the HIPASS/ZOA data, scaled
accordingly and removed from each spectrum after bandpass
correction. The Sun and many other strong continuum sources 
in the Galactic plane when seen in the beam sidelobes may generate {\em
  off-axis ripple}. This ripple has no characteristic signature, is often quite
complex and may vary rapidly. \citet{Calabretta:2014} describe a
method to filter this ripple, mainly being of use for point
sources. It was not applied in the present RRL data reduction in order
to avoid possible side effects on the extended emission.

After bandpass correction and calibration, the three RRLs are
extracted from each 1024 channel spectrum, shifted by the correct
fractional number of channels by Fourier interpolation and then
stacked. The errors introduced by stacking the lines in
  frequency space rather than in velocity space are less than 2 per cent of the
channel width. The three lines are stacked using a weighted mean, where the weights
are proportional to the square root of the bandpass response measured
at each line's rest frequency. The weights are 0.35, 0.39 and 0.26 for
the H168$\alpha$, H167$\alpha$ and H166$\alpha$; the lower
H166$\alpha$ weight is due to the Gibbs ringing close to the strong
\hi~line (figure 2 in \pap). Averaging of the RRLs improves the signal-to-noise of
the final line by a factor of $\sim \sqrt 3$. The number of channels extracted was set to 51, the
maximum number accommodating the proximity of the H166$\alpha$ line to the
edge of the band, and is equivalent to an LSR (local standard of rest)
velocity range of $\pm 335$\kms.

The stacked RRL spectra are then gridded using
\textit{gridzilla}\footnote{Provided in the same package as \textit{livedata}.}
\citep{Barnes:2001} into 4\,arcmin pixels. The spectrum at each pixel
is generated from the weighted median of the stacked spectra within
a cut-off radius of 6\,arcmin. The weight for each input spectrum is based
on its angular distance from the pixel. Even though the weighted
median estimation is non-linear, it is robust against RFI (radio
frequency interference) and other
time-variable sources of emission such as the Sun, and it also
produces a gridded beam that is very close to Gaussian with
14.4\,arcmin FWHM. However, one effect of median
gridding is that the flux density scale differs depending on the
source size (\paper). This problem has now been overcome by iterative gridding
\citep{Calabretta:2014} using a gain factor of 1.2 in order to
minimize the number of iterations and consequent increase in the rms
noise. Simulations indicate that a gain factor of 1.2 recovers the
peak height and integrated flux of sources of various sizes within
1 per cent, with a negligible effect in the final beam FWHM.

This procedure is applied to the ZOA and the HIPASS datasets
separately, which are combined by correcting
the ZOA data for a latitude-dependent offset and averaging both
datasets, giving five times more weight to the ZOA survey (\paper). The data
are further Tukey smoothed, hence the spectral resolution is degraded
from 16\kms~to 20\kms. 

\subsection{Calibration}
\label{sec:datacal}

\begin{figure}
\centering
\includegraphics[scale=0.45]{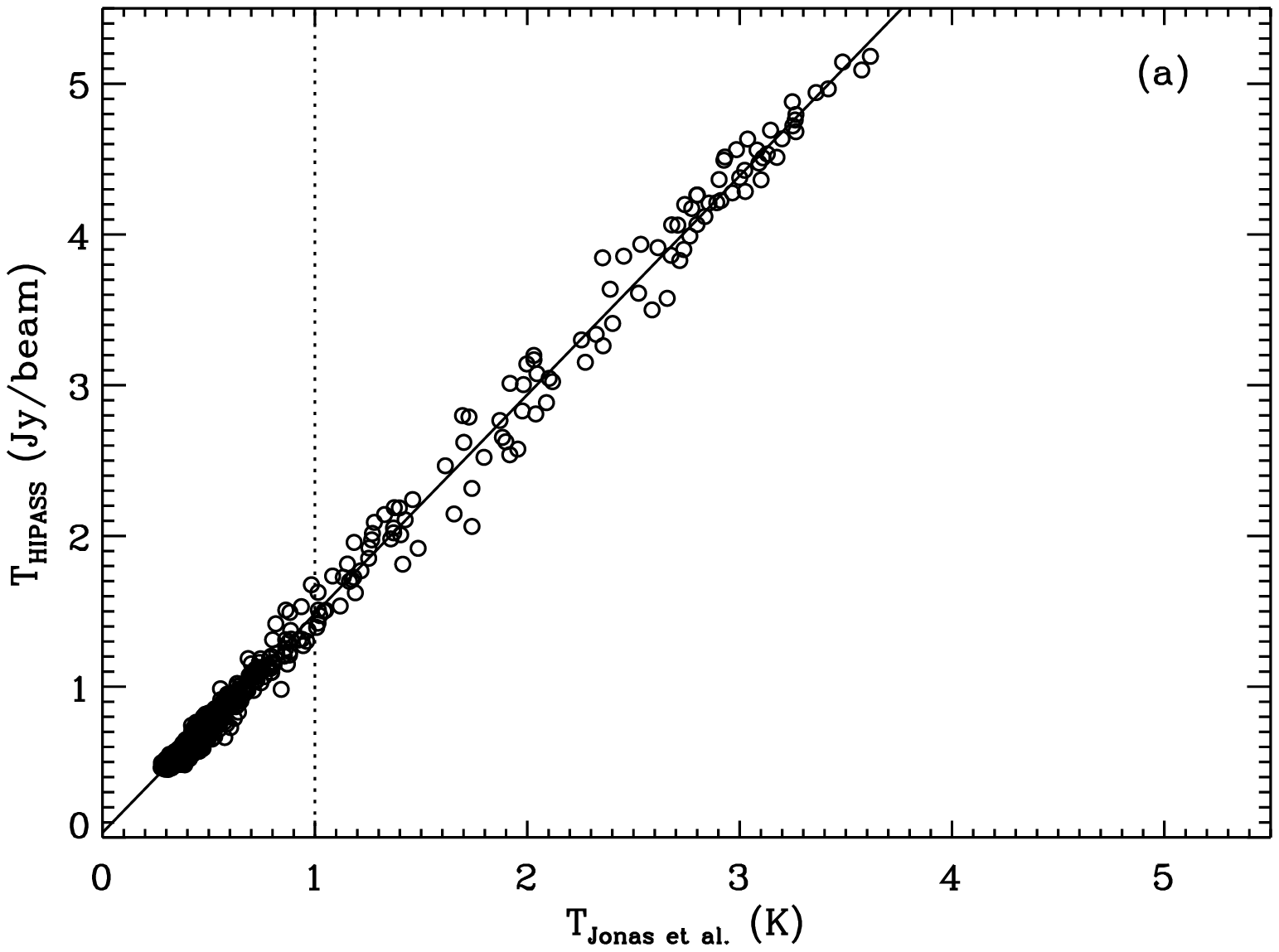}
\includegraphics[scale=0.45]{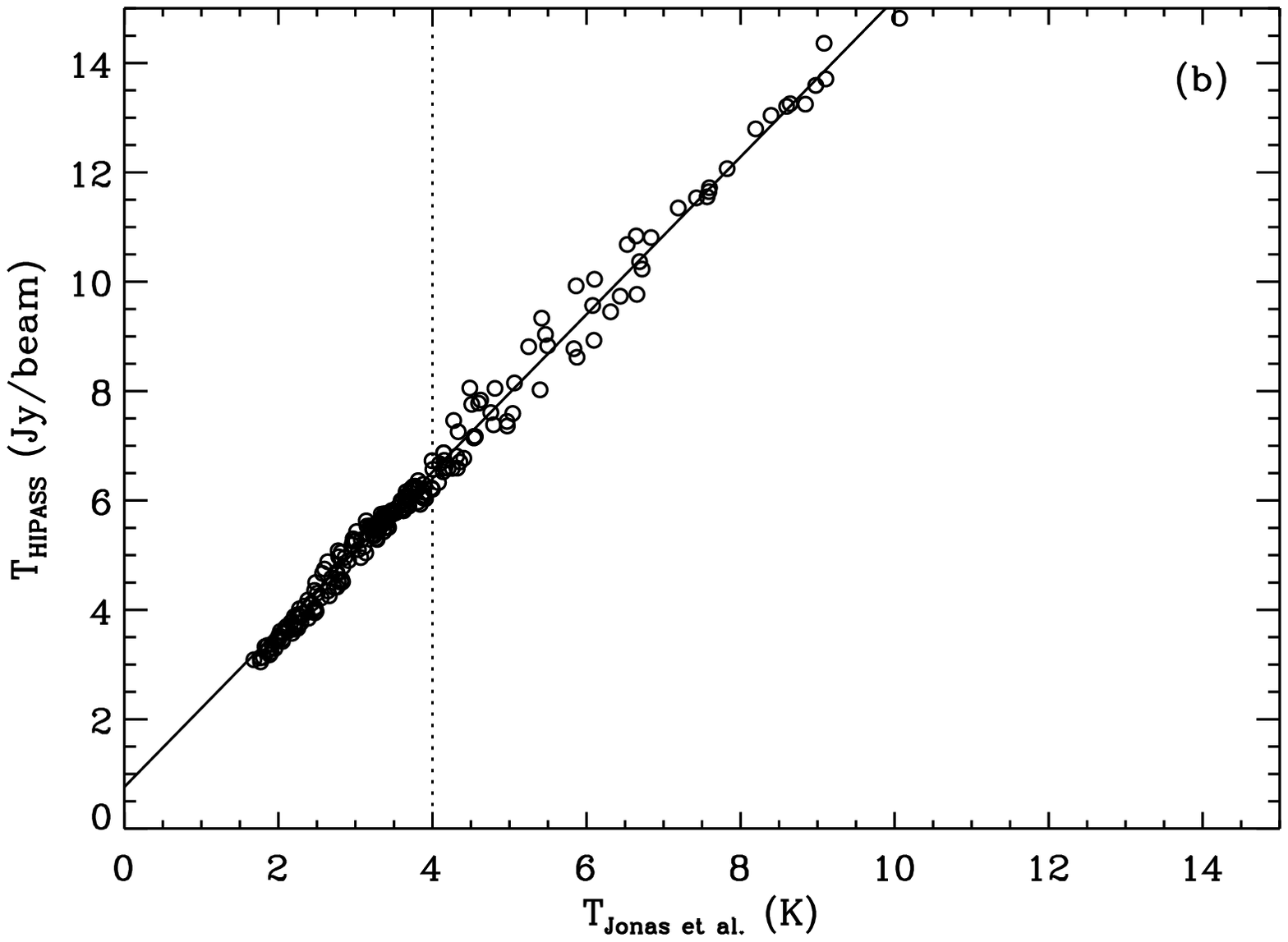}
\includegraphics[scale=0.45]{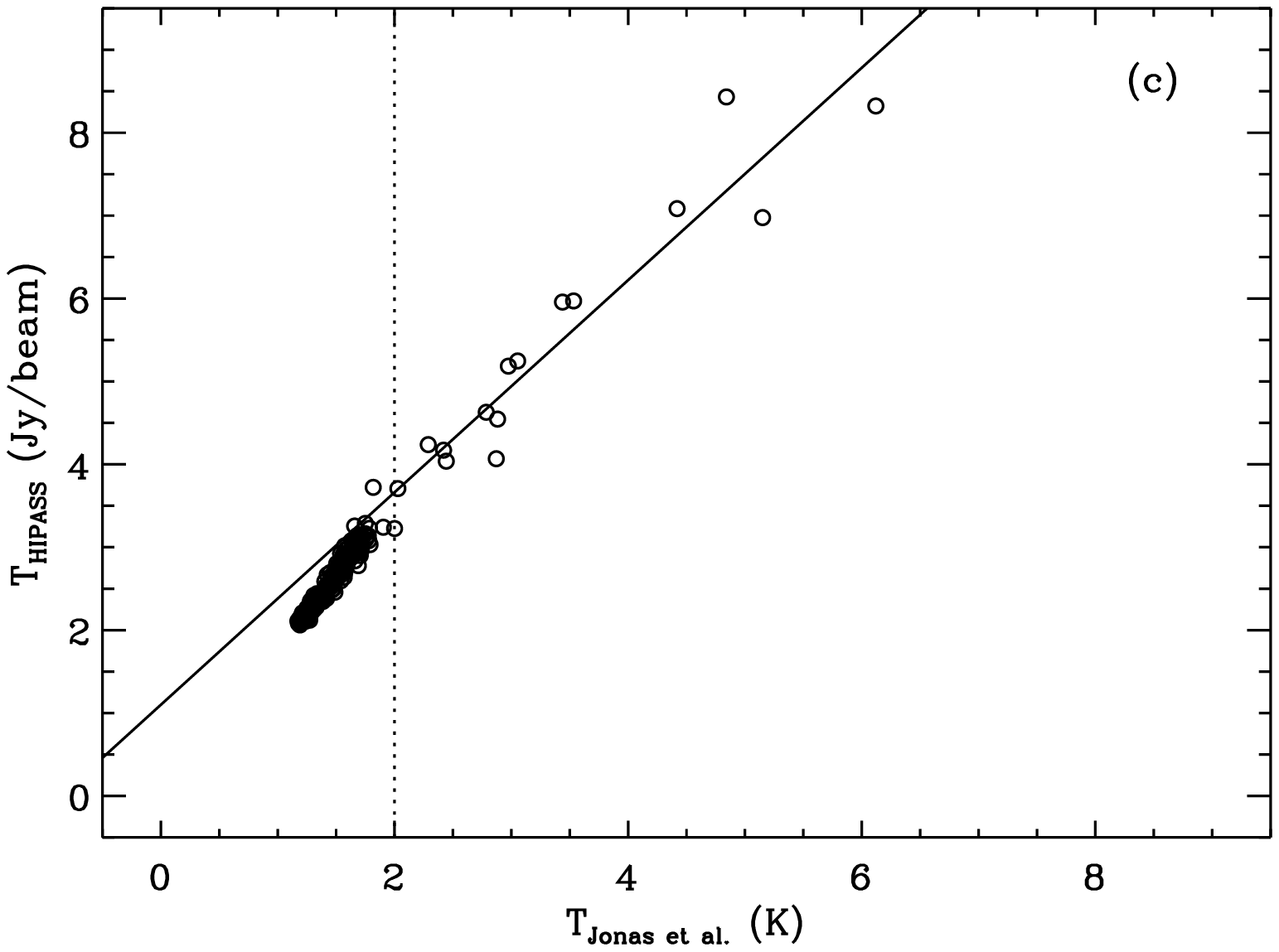}
\caption{T-T plots for (a) the Rosette nebula, (b) W35, and (c) W40, showing
  the correlation between the HIPASS/ZOA and the \citet{Jonas:1998}
  surveys. The HIPASS/ZOA data have been smoothed to 20\,arcmin
  resolution and extrapolated from 1.4 to 2.3\,GHz. The vertical dotted lines
  indicate the values above which the linear fit, shown by the
  solid lines, is performed. \label{fig:cal1}}
\end{figure}


\begin{table*}
\centering
\caption{The first 10 entries of the list of 317 \hii~regions in the
range $l=196\degr$ --
$0\degr$ -- $52\degr$ and $|b| \leq 5\degr$,
extracted from the thermal continuum map of Fig. \ref{fig:intmap} using SE{\sc
 xtractor}. The full catalogue is available online. Column 1 numbers each object; columns 2 and 3 are the
Galactic coordinates; columns 4 and 5 give the angular size of the
source, where $\theta_{a}$ and $\theta_{b}$ are the measured values $a$
and $b$ deconvolved with the 14.4\,arcmin beam; $p$ indicates that the
object is less than 10\,arcmin in size; column 6 is the
PA measured counterclockwise between the semi-major axis of the
source and the longitude axis; columns 7 and 8 give the peak and total flux
density at 1.4~GHz; column 9 gives the RRL central velocity from a one- or
two-component Gaussian fit; column 10 gives the internal flags generated by SE{\sc
 xtractor}$^{\rm a}$; column 11 gives general remarks on the
source$^{\rm b}$.
 \label{tab:hiicat}}
\begin{tabular}{c|cccccccccc}
\hline
\hline
Number & $l$ & $b$  & $\theta_{a}$ & $\theta_{b}$ & PA & $S_{\rm p}$ &
$S$ & $V$ & Flag $^{\rm a}$ & Notes$^{\rm b}$ \\
 & ($\degr$) & ($\degr$) & (arcmin) & (arcmin) & ($\degr$) &
 (Jy\,beam$^{-1}$) & (Jy) &  (km\,s$^{-1}$)& & \\
\hline
$  1$&$  1.14$&$  0.01$&$ 14.9$&$p$&$-45$&$  20.4\pm   4.0$&$  29.3\pm   5.8$&$ -20.5$&$0$&                           C  [1]\\
$  2$&$  2.21$&$  0.34$&$p$&$p$&---&$   8.0\pm   1.6$&$   8.0\pm   1.6$&$   4.7$&$2$&---\\
$  3$&$  2.44$&$  0.17$&$ 17.2$&$p$&$-19$&$   8.0\pm   1.7$&$  12.4\pm   3.5$&$   4.3$&$2$&---\\
$  4$&$  3.18$&$ -0.03$&$ 17.2$&$ 14.5$&$  5$&$  11.7\pm   2.9$&$  25.8\pm   6.9$&$   4.0$&$0$&---\\
$  5$&$  4.33$&$  0.20$&$ 13.6$&$p$&$  7$&$  12.3\pm   2.8$&$  16.9\pm   4.2$&$   4.4$&$0$&                             [1]\\
$  6$&$  4.47$&$ -0.11$&$p$&$p$&---&$   9.7\pm   2.7$&$   9.7\pm   2.7$&$   9.6$&$0$&                             [1]\\
$  7$&$  4.73$&$  0.17$&$p$&$p$&---&$   8.4\pm   2.5$&$   8.4\pm   2.5$&$   7.4$&$0$&---\\
$  8$&$  5.18$&$ -0.16$&$p$&$p$&---&$   5.7\pm   2.1$&$   5.7\pm   2.1$&$   5.6$&$0$&---\\
$  9$&$  5.24$&$  0.17$&$ 14.4$&$p$&$ 11$&$   6.6\pm   2.5$&$   9.3\pm   4.0$&$   9.7$&$2$&                            [1]\\
$ 10$&$  5.84$&$ -0.37$&$ 12.6$&$p$&$  4$&$  33.1\pm   5.9$&$  44.0\pm   8.0$&$  12.5$&$2$&                        W28 [1]\\
\hline
\end{tabular}
\\
$^{\rm a}$ Flag parameter: 2 - the source was originally blended
with another one; 8 - the source is truncated owing to its proximity
to the image boundary. $^{\rm b}$  These are taken from \citet{Paladini:2003} and follow the
definitions given by \citet{Kuchar:1997} to define if a source is: C,
in a complex field; S, with a strong source nearby; X, with a strong
source nearby ($<10$\,Jy), radio or optical counterpart. The value
within the square brackets corresponds to the number of sources in the
\citeauthor{Paladini:2003} catalogue that match a given source in this survey.
\end{table*}

During observations, the system temperature was calibrated against a
high-quality noise diode switched in and out of the signal path. The
diode itself was calibrated periodically against flux density
calibrators. Gaussian fits to the peak heights of the source 1934-638 (14.9\,Jy at 1420\,MHz) and Hydra A
(40.6\,Jy at 1395\,MHz), resulted in a mean calibration factor of 1.09 (1.11 and
1.07 for 1934-638 and Hydra A respectively;
\citealt{Calabretta:2014}).

A detailed knowledge of the beam pattern and of the sky brightness
distribution is required in order to convert the data correctly to absolute
temperature units in the presence of both compact and extended
emission (e.g., \citealt{Rohlfs:2000}). However, both these quantities are difficult to
measure and thus the conversion is usually performed in the full beam
scale, using a single value across the whole map. The full beam
corresponds to the integral of the observed beam pattern out to a large
radius, whereas the main beam is measured within the first sidelobes \citep{Jonas:1998}.

In \paper\ we converted the data to
the main beam scale following the Rayleigh-Jeans relation for a
Gaussian FWHM of 14.4\,arcmin at the frequency of the H167$\alpha$
line, obtaining a value of $T_{\rm
  b}/S=0.84$\,K\,/(Jy\,beam$^{-1}$). This scale is appropriate for
compact sources and
overestimates the emission from sources more extended than the
beam. The above value compares with 0.80 \,K\,/(Jy\,beam$^{-1}$), used in
\hi~observations of extended objects with the Parkes telescope
\citep{Staveley-Smith:2003}, and which is appropriate for the somewhat
higher mean efficiency of the inner seven beams. A lower value of about 0.57\,K\,/(Jy\,beam$^{-1}$)
is expected for the outer beams. 
An alternative to the
use of theoretical conversion factors is the comparison with other
data sets. \citet{Calabretta:2014} converted the 1.4\,GHz continuum map by
correlating the HIPASS/ZOA data with an existing map at the same
frequency but lower resolution, 35.4\,arcmin, and which is calibrated in
the full beam brightness scale for a very extended beam of $7\degr$ \citep{Reich:1982,
  Reich:1986, Reich:2001}. The authors derive an empirical factor of
0.44\,K\,/(Jy\,beam$^{-1}$), 30 per cent lower than the theoretical value of
0.57\,K\,/(Jy\,beam$^{-1}$). This discrepancy is likely due
to power on intermediate and large scales ($\gtsim 7\degr$),
dominating the cross-plot of the HIPASS/ZOA versus
the Reich et al. data used to derive the conversion factor.

In this work we perform an independent analysis by comparing the
HIPASS/ZOA continuum data with that from the
2.3\,GHz survey by \citet{Jonas:1998}, at 20\,arcmin resolution. 
The authors quote an uncertainty in the full beam scale of less than
5 per cent, for a beam efficiency of 69 per cent.
Even if the two surveys cover a similar area of the sky, we need to
perform the comparison over restricted regions, since the datasets have
to be extrapolated to the same frequency using a fixed spectral index.
We focus on regions dominated by thermal emission, thus located in the
Galactic plane, and within the limits of the RRL survey, selecting
three \hii~regions: the Rosette nebula 
(G206.2-2.1), W35 (G18.5+2.0), and W40 (G28.8+3.5). These are three of
the few objects in the area under study that are well isolated and
outside the Galactic plane (see Fig.~\ref{fig:intmap}).

\begin{figure*}
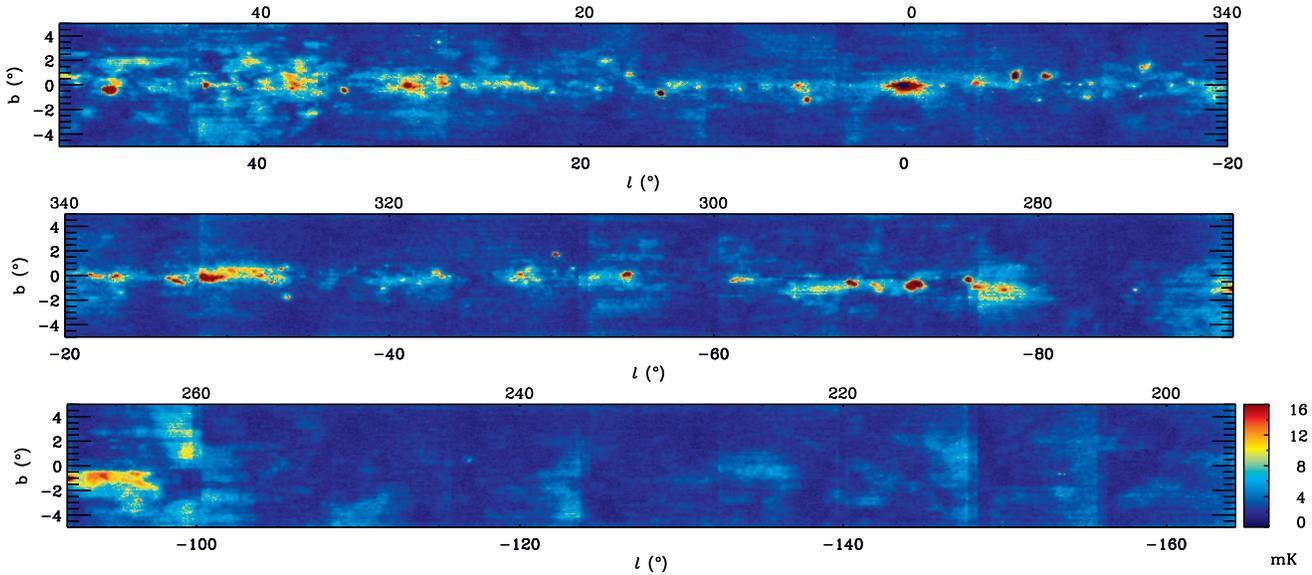

\hspace*{-0.95cm}\includegraphics[scale=0.92,angle=90]{Figs/fig2a.epsi}
\hspace*{-1.0cm}\includegraphics[scale=0.92,angle=90]{Figs/fig2b.epsi}
\includegraphics[scale=0.92,angle=90]{Figs/fig2c.epsi}
\caption{Map of the rms noise per channel, estimated using the
  emission-free ends of each spectrum (see text). The longitude coverage of the survey is divided into
   three sections: $l=340\degr$ -- $0\degr$ -- $52\degr$,
   $l=268\degr$ -- $340\degr$, and $l=196\degr$ -- $268\degr$. The typical
   rms noise in the Galactic plane is 4.5\,mK, increasing by nearly a factor of 10 towards
   some of the brightest \hii~regions. Note the
  two longitude labels above and below each panel. \label{fig:rmsmap}}
\end{figure*}

\begin{figure*}
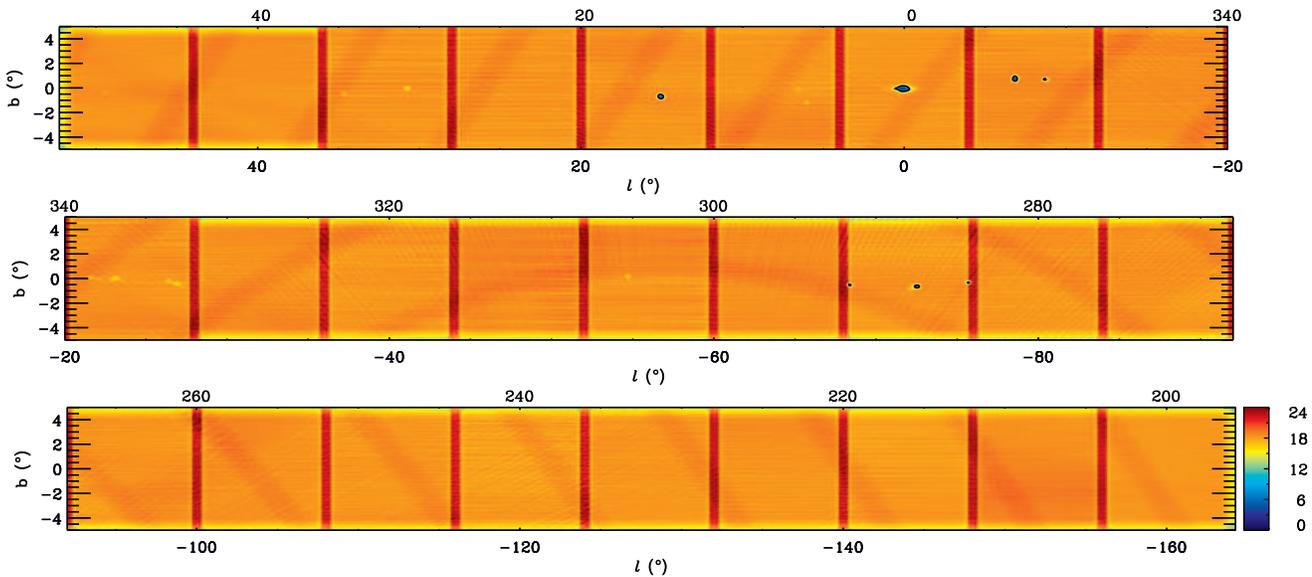

\hspace*{-0.95cm}\includegraphics[scale=0.92,angle=90]{Figs/fig3a.epsi}
\hspace*{-1.0cm}\includegraphics[scale=0.92,angle=90]{Figs/fig3b.epsi}
\includegraphics[scale=0.92,angle=90]{Figs/fig3c.epsi}
\caption{Beam coverage map given by the sum in quadrature of beam
  weights used to produce the final data cube and maps. One unit
  corresponds to one independent boresight observation. The longitude coverage of the survey is divided into
   three sections as in Fig. \ref{fig:rmsmap}. Typical values
  are around 18, increasing to about 22 where ZOA zones overlap. The
  saturated pixels have values below 10
  (within the black contour). This map illustrates the striations along the
  HIPASS and ZOA scan directions, as well as the overlap regions
  between the different zones. \label{fig:sensmap}} 
\end{figure*}

Fig.~\ref{fig:cal1} shows the resulting T-T
plots. The HIPASS/ZOA continuum emission is extrapolated from 1.4 to
2.3\,GHz using a spectral index $\beta=-2.1$, where $T\propto
\nu^{\beta}$ \citep{D3:2003,Draine:2011}, and smoothed from 14.4 to
20\,arcmin resolution. We select regions of $4\degr\times 4\degr$
centred on the Rosette and
$2\degr\times 2\degr$ centred on W35 and W40. The maps are
gridded to a common pixel size of 8\,arcmin and all of the pixels for each region are shown in Fig.~\ref{fig:cal1}.
The correlation between the two data sets gives
conversion factors of $0.68\pm0.10$, $0.69\pm0.05$, and $0.72\pm0.08$\,K\,/(Jy\,beam$^{-1}$)
for the Rosette, W35, and W40 \hii~regions respectively, where the
uncertainties correspond to the scatter of the points relative to the
best fit line. The correlation is performed using the points above a given threshold value in
temperature, shown by the vertical dotted line in each panel of
Fig. \ref{fig:cal1}. This threshold is applied in order to exclude
background emission, which also includes synchrotron emission at these
frequencies and latitudes (\paper). The background level is estimated by fitting a
Gaussian profile with a constant term to the 2.3\,GHz maps. The
increase in the conversion factor from one object to
the other is likely related to their angular size: W40 has a
FWHM of about $\sim10$\,arcmin compared to $\sim 40$ and $\sim60$\,arcmin
for the W35 and Rosette \hii~regions, respectively (Table~\ref{tab:hiicat}). The Rosette and W35 \hii~regions are
representative of emission broader than the 14.4\,arcmin beam and
extended over scales of $\ltsim 1\degr$, whereas W40 has a size comparable to
the beam, hence a factor closer to 0.80\,K\,/(Jy\,beam$^{-1}$). The
estimated conversion factors are consistent within their
uncertainties, with a mean value of $T_{\rm
  b}/S=0.70\pm0.03$\,K\,/(Jy\,beam$^{-1}$).
If we vary the background levels by 1\,K, the mean
calibration factor changes by 0.02\,K\,/(Jy\,beam$^{-1}$); this difference
is included in the 4 per cent uncertainty on $T_{\rm b}/S$. 

We use the above value of 0.70\,K\,/(Jy\,beam$^{-1}$) to convert the RRL data into
brightness temperature, given that
the instrument and data are the same and that 
the reduction of the HIPASS/ZOA continuum data is
similar to that of the
spectral lines. We can also estimate the conversion factor
using the free-free map derived from the RRL integrated intensity
(using Eq. (\ref{eq:ff}) and described in Section
\ref{subsec:ff}) as we did with the HIPASS/ZOA total continuum. The
mean conversion factor obtained from the analysis of the same three
regions is $T_{\rm b}/S = 0.68\pm0.11$\,K\,/(Jy\,beam$^{-1}$), which is consistent with
that derived above using the total continuum. The higher uncertainty
results partly from the larger scatter of the points, associated with
a higher noise level, but also from the uncertainty on the electron
temperature, which will be discussed in Section \ref{subsec:ff}.

Finally, we adopt $T_{\rm b}/S = 0.70$\,K\,/(Jy\,beam$^{-1}$) along
with a conservative calibration uncertainty
of 10 per cent, considering that both the 1.09 factor above and the
brightness temperature conversion are based on only two and three
sources respectively. We note that for a direct comparison of the
RRL data with the HIPASS/ZOA continuum data, the latter need to be
corrected by a factor of 1.6, the ratio between the conversion factor
found here (0.70 \,K\,/(Jy\,beam$^{-1}$)) and that derived in
\citet{Calabretta:2014} (0.44 \,K\,/(Jy\,beam$^{-1}$)).

\subsection{Final data cube}
\label{sec:effects}

The final $216\degr \times 10\degr \times 670$\kms\ data cube covers
the $l$-range $196\degr$ -- $0\degr$ -- $52\degr$, $|b| \leq 5\degr$
and $V_{\rm LSR}=\pm 335$\kms. Fig.~\ref{fig:rmsmap} shows the
spectral rms noise map, for the velocity resolution of 20\kms. We
estimate the standard deviation of each spectrum using 
its emission-free ends, which correspond to 38 out of the 51 channels once we
allow for the RRL emission to spread within the conservative range of
$\pm85$\kms\ about its peak velocity (Section \ref{subsec:rrlint}). 
The central RRL velocity is found using a three-point quadratic fit around the brightest
peak ({\sc miriad}\footnote{www.atnf.csiro.au/computing/software/miriad.html}
task {\sc moment}, \citealt{Sault:1995}). The rms noise per
channel in the cleanest regions of the survey is 2.8\,mK. However, the
typical value in the Galactic plane, as well as close to the regions
where individual $8\degr (l) \times 10\degr (b)$ ZOA scans overlap, is
4.5\,mK or 6.4\,mJy\,beam$^{-1}$. At the location of some of the
brighest continuum sources, the spectral noise reaches 
20 -- 40\,mK; this is due to the stronger baseline ripple effect and
spectral ringing from Galactic \hi\ emission. 
Nevertheless, the regions of highest rms noise also have high ($>3$)
signal-to-noise spectra.

The vertical discontinuities visible in the
spectral rms maps are also seen in some of the RRL channel
maps that are presented in Section \ref{subsec:rrlmaps}
(Fig. \ref{fig:cmaps1}). The overlap of individual scans is clearer in the beam coverage map of
Fig. \ref{fig:sensmap}, which shows the square
root of the sum of squares of the beam weights used in producing the
spectral cube. One unit in this map
  corresponds to one independent boresight observation. Over these
regions and especially outside the Galactic plane,
some of the channel maps show negative values. This is likely due to
differences in the baseline levels, related to the bandpass correction
(Section \ref{sec:datanalaysis}). This effect is, nevertheless, at a
level of about 2 times the rms noise. Striations are seen along the direction
of the HIPASS scans, which cross the Galactic plane at various
angles, in some of the velocity channels and zones, e.g., $l
\sim 320\degr$ and $V=26.8$\kms, or $l
\sim 270\degr$ and $V=-80.3$\kms\ (Fig. \ref{fig:cmaps1}). This is a
low level effect, of the order of
$\sim$1 -- 2\,mK. 

\begin{figure}
\centering
\includegraphics[scale=0.45]{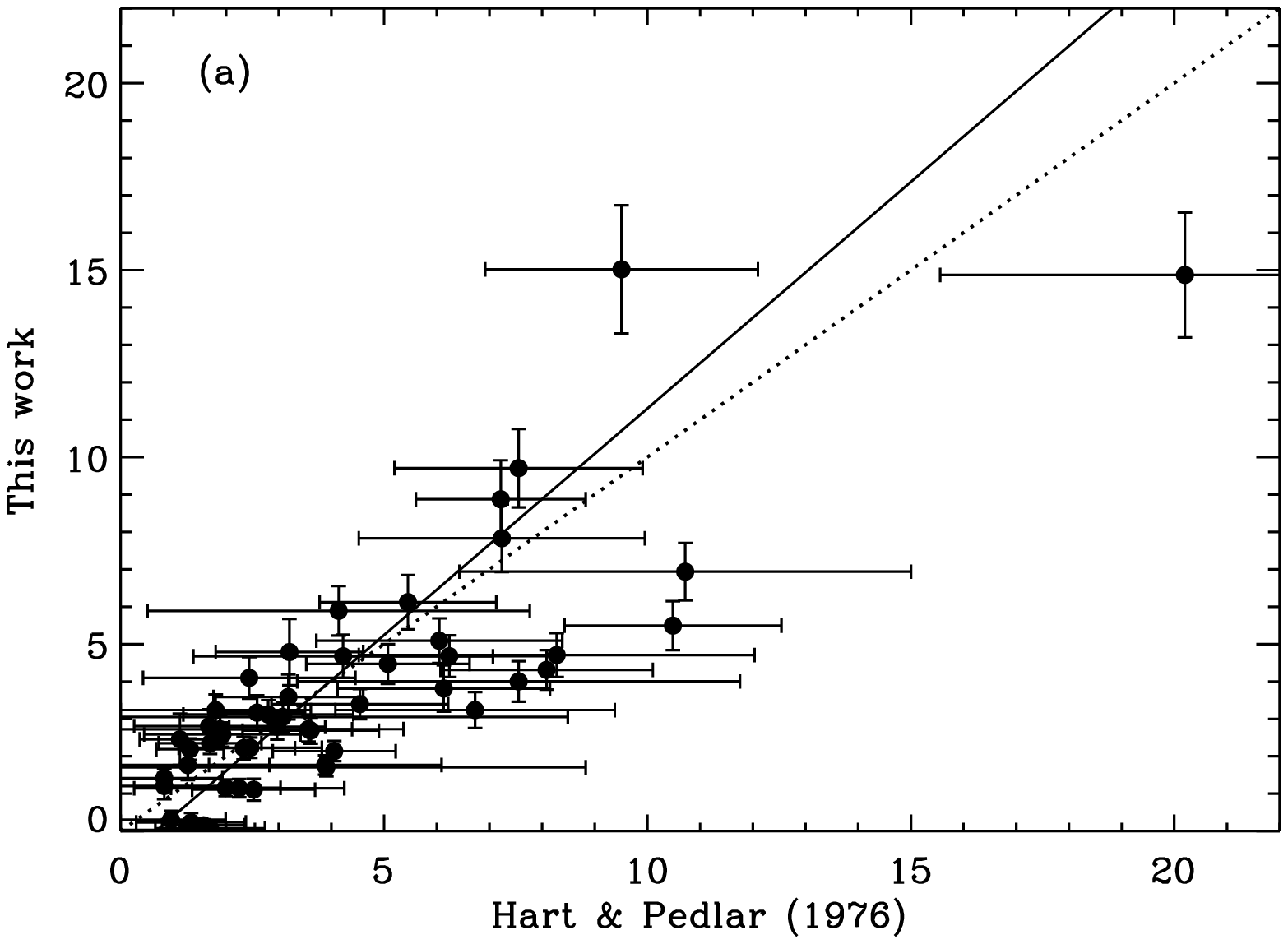}
\includegraphics[scale=0.45]{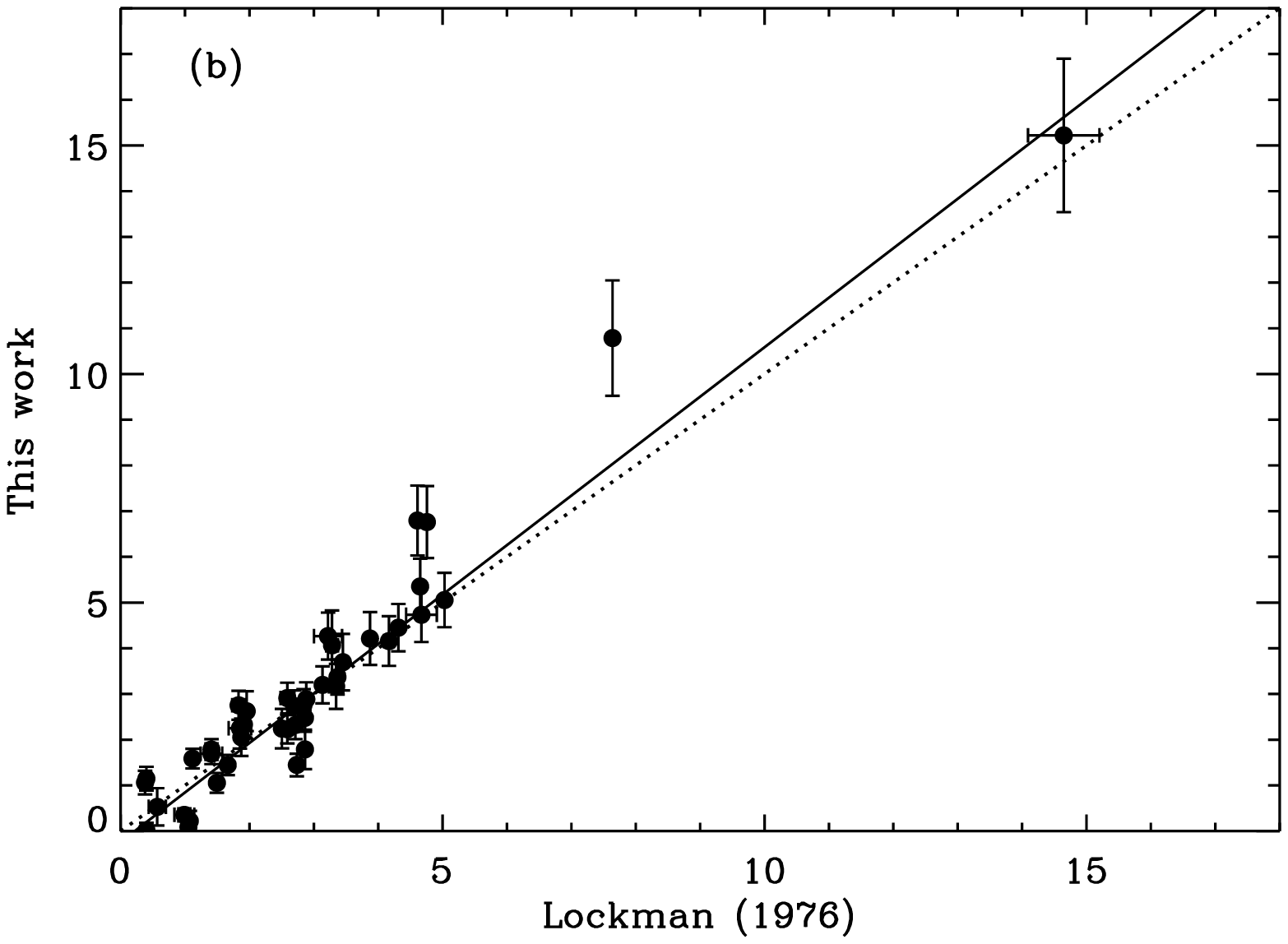}
\includegraphics[scale=0.45]{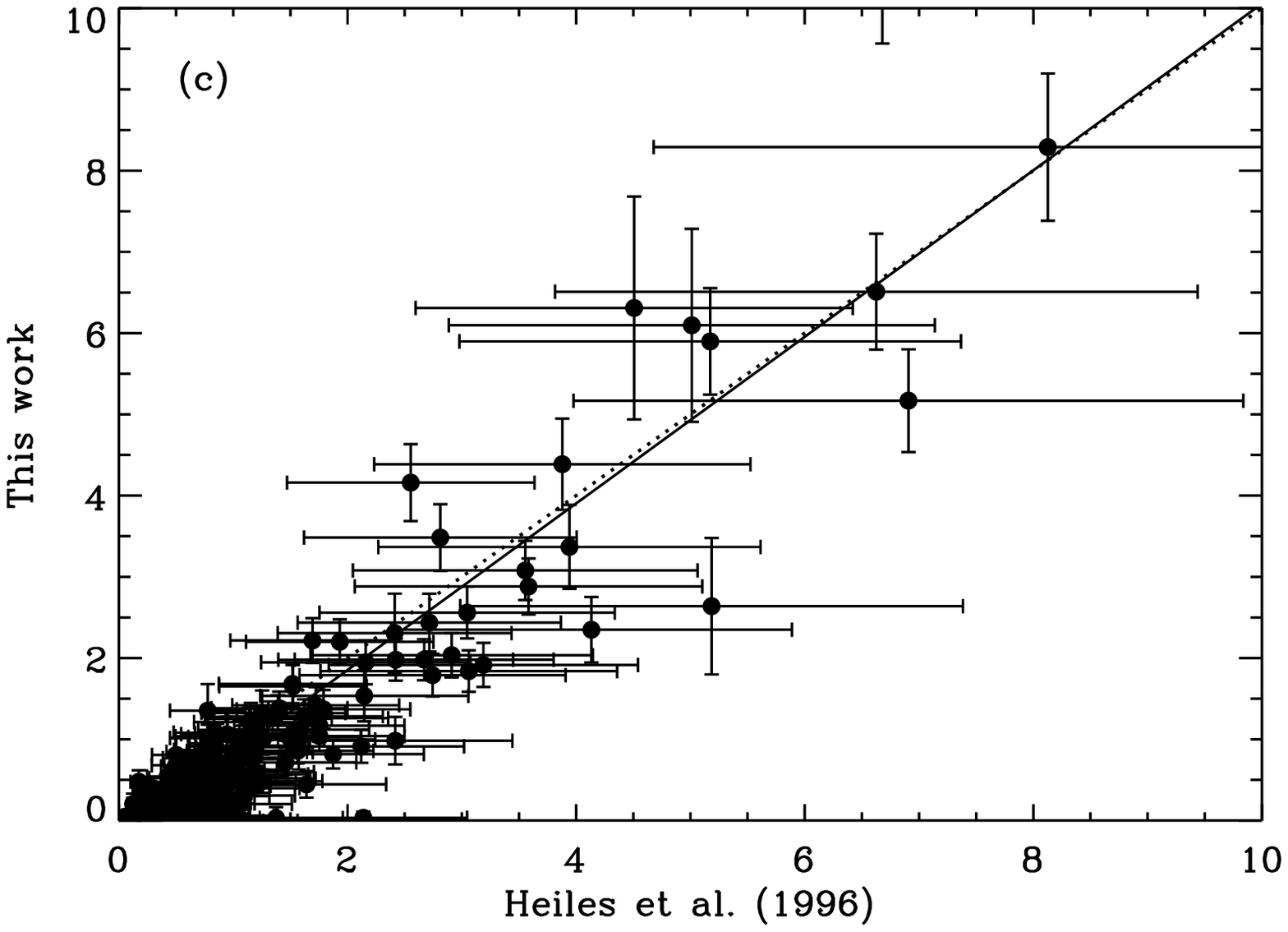}
\caption{Comparison between the HIPASS/ZOA RRL data and those from the
  surveys of (a) \citet{Hart:1976}, (b) \citet{Lockman:1976}, and
  (c) \citet{Heiles:1996}. The quantity plotted, with the
  corresponding error bars, is the RRL integrated intensity in units
  of K\,\kms. The dashed lines indicate equality and the solid lines
  correspond to the linear fit between the different data sets. There
  are 51, 43, and 208 points in panels (a), (b), and (c),
  respectively. \label{fig:cal2}}
\end{figure}

Some of the pixels in the maps are
blanked as the corresponding data are affected by sources strong
enough to saturate the receiver electronics. There are 129 pixels
affected, seen in the beam coverage map of Fig. \ref{fig:sensmap} with
values below 10 (e.g., 63 pixels in the Galactic centre, 21 pixels in
M17 G15.1-0.7).

The spectral negatives, which originate from
the bandpass distortions (Section~\ref{sec:datanalaysis}), are seen at the
position of some strong sources in Fig.~\ref{fig:cmaps1}. They appear in the velocity channels
adjacent to those corresponding to the peak emission of the
object. For example, the \hii~region M8 G6.0-1.2 peaks at
$V\sim0$\kms\ and a negative of about $10$ per cent of its peak temperature is seen at
the same position, at $V=53.6$\kms. Stronger spectral negatives are produced by an interfering signal at
$V \approx -130$\kms, which significantly affects a large part of the data,
especially in the fourth quadrant of the Galactic plane. In this
longitude range, $l\sim 270\degr$ -- $330\degr$, the interfering signal
is at a similar velocity to that expected from helium and carbon RRLs
(Section~\ref{sec:hecrrls}). However this effect is small when observing low in the north (i.e., at
$l\sim30\deg$). The effects of the interference can be seen around $l=312\degr$ at $V=-147.3$\kms\ and
$l=288\degr$ at $V=-133.8$\kms\ (visible in the longitude-velocity
diagram of Fig.~\ref{fig:lonvel}). The former is probably associated with
the radio galaxy Centaurus B, $(l,b)=(309.7\degr,1.7\degr)$,
where a negative spike of about 200\,mK is detected. The same negatives
are present in the spectra of neighbouring \hii~regions, nevertheless at $\sim
90$\kms\ away from their emission peak, which is at $V\sim-50$\kms. The spectral negatives at
$l\sim 288\degr$ are seen mostly around the three close-by bright
\hii~regions that have saturated pixels (Fig.~\ref{fig:cmaps1}). Similar to the previous case, this distortion of
the spectra by strong continuum sources does not seem to affect the peak of
the hydrogen lines that are about 100\kms\ away. This can, however,
affect the helium and carbon spectra (Section~\ref{sec:hecrrls}).

In summary, the data show some striations and discontinuities,
which are large-scale artifacts close to the noise level, as
well as more significant defects, the spectral negatives that
originate from bandpass distortions. Nevertheless, as will be shown
in the next section, such artifacts do not affect the quality of the
data within the assumed calibration uncertainty of 10 per cent.

\subsection{Comparison with previous RRL observations}
\label{sec:rrlcomp}

The accuracy of the present data can be checked by comparison
with previous RRL observations. We consider three
surveys made at a similar frequency and with comparable angular
resolutions, but with different telescopes and observing
techniques. \citet{Hart:1976} and
\citet{Lockman:1976}, hereafter HP76 and L76 respectively, measured the
H166$\alpha$ line at positions along the first quadrant of the
Galactic plane. HP76 used the Mk II telescope, with a beam
FWHM of $31\times33$\,arcmin, whereas the L76 observations
were performed at an angular resolution of
21\,arcmin with the 43\,m NRAO telescope. \citet{Heiles:1996},
hereafter HRK96, targeted hundreds of positions mostly towards the inner Galaxy, but restricted
to $|b|> 0\pdeg6$, with the
NRAO and the 26\,m HCRO telescope with 36\,arcmin beam. We use the
data from the HCRO telescope,
which observed most of the positions primarily at the H165$\alpha$ and
H167$\alpha$ lines. 

\begin{figure*}
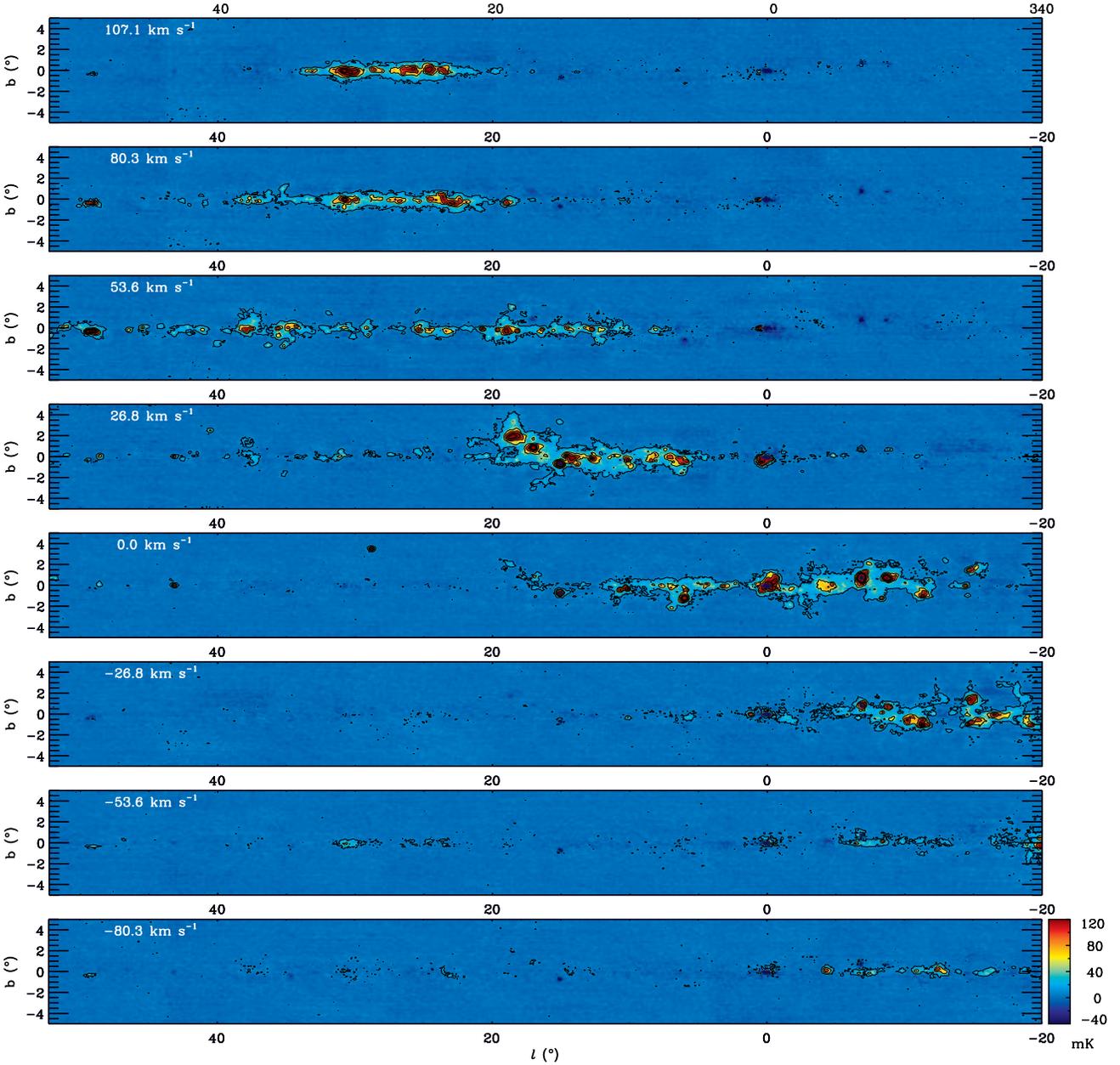

\hspace*{-0.8cm}\includegraphics[scale=0.92,angle=90]{Figs/fig5a.epsi}
\hspace*{-0.8cm}\includegraphics[scale=0.92,angle=90]{Figs/fig5b.epsi}
\hspace*{-0.8cm}\includegraphics[scale=0.92,angle=90]{Figs/fig5c.epsi}
\hspace*{-0.8cm}\includegraphics[scale=0.92,angle=90]{Figs/fig5d.epsi}
\hspace*{-0.8cm}\includegraphics[scale=0.92,angle=90]{Figs/fig5e.epsi}
\hspace*{-0.8cm}\includegraphics[scale=0.92,angle=90]{Figs/fig5f.epsi}
\hspace*{-0.8cm}\includegraphics[scale=0.92,angle=90]{Figs/fig5g.epsi}
\includegraphics[scale=0.92,angle=90]{Figs/fig5h.epsi}
\caption{Maps of the RRL emission at different velocity
  channels, as indicated in the top left corner of each
   panel. The longitude coverage of the survey is divided into
   three sections as in Fig. \ref{fig:rmsmap}. Contours
    are at 10, 50, 90, 190, 290, and 390\,mK. \label{fig:cmaps1}} 
\end{figure*}

The results are shown in Fig. \ref{fig:cal2},
where we compare the RRL integrated intensities in order to take into account the different velocity
resolutions of the surveys. The HIPASS data are smoothed to the lower
angular resolutions of 32, 21, and 36\,arcmin to match the observing
beams of HP76, L76, and HRK96, respectively;  the data are further regridded to pixel sizes
of 9, 6, and 10\,arcmin.  The uncertainties in the HIPASS
intensities follow from the propagation of the spectral noise
(Fig. \ref{fig:rmsmap}) to the line integral, as described in
Section \ref{subsec:rrlint}. We also add in quadrature the 10 per cent systematic
uncertainty, associated with the brightness temperature calibration (Section
\ref{sec:datacal}). The correlation between the HIPASS and
the three different surveys gives linear slopes of $1.2\pm0.2$,
$1.08\pm0.04$, and $1.03\pm0.06$, for HP76, L76, and HRK96,
respectively. The linear fits take into account the uncertainties in
both abscissa and ordinate\footnote{We use only the upper error limits
  given by HP76; the uncertainties listed in L76 are divided by 3 to
  obtain 1$\sigma$ estimates; we assume a conservative 30 per cent
  fractional error on both the line temperatures and widths fitted by
  HRK96.}. These results indicate that the present
RRL data are consistent with previous observations at the 10 per
cent level, and thus give confidence in the overall accuracy and
calibration of the current survey.

\section{The ionized Galactic disk}
\label{sec:fftemp}

In this section we present the RRL emission at 1.4\,GHz in the plane
of the Galaxy and discuss its three-dimensional distribution. 

\subsection{RRL maps}
\label{subsec:rrlmaps}

\begin{figure*}
  \ContinuedFloat
\hspace*{-1cm}\includegraphics[scale=0.92,angle=90]{Figs/fig5i.epsi}
\hspace*{-1cm}\includegraphics[scale=0.92,angle=90]{Figs/fig5j.epsi}
\hspace*{-1cm}\includegraphics[scale=0.92,angle=90]{Figs/fig5k.epsi}
\hspace*{-1cm}\includegraphics[scale=0.92,angle=90]{Figs/fig5l.epsi}
\hspace*{-1cm}\includegraphics[scale=0.92,angle=90]{Figs/fig5m.epsi}
\hspace*{-1cm}\includegraphics[scale=0.92,angle=90]{Figs/fig5n.epsi}
\hspace*{-1cm}\includegraphics[scale=0.92,angle=90]{Figs/fig5o.epsi}
\includegraphics[scale=0.92,angle=90]{Figs/fig5p.epsi}
\caption{ \label{fig:cmaps1}} 
\end{figure*}

\begin{figure*}
  \ContinuedFloat
\hspace*{-1.1cm}\includegraphics[scale=0.92,angle=90]{Figs/fig5q.epsi}
\hspace*{-1.1cm}\includegraphics[scale=0.92,angle=90]{Figs/fig5r.epsi}
\hspace*{-1.1cm}\includegraphics[scale=0.92,angle=90]{Figs/fig5s.epsi}
\hspace*{-1.1cm}\includegraphics[scale=0.92,angle=90]{Figs/fig5t.epsi}
\hspace*{-1.1cm}\includegraphics[scale=0.92,angle=90]{Figs/fig5u.epsi}
\hspace*{-1.1cm}\includegraphics[scale=0.92,angle=90]{Figs/fig5v.epsi}
\hspace*{-1.1cm}\includegraphics[scale=0.92,angle=90]{Figs/fig5x.epsi}
\includegraphics[scale=0.92,angle=90]{Figs/fig5z.epsi}
\caption{ \label{fig:cmaps1}} 
\end{figure*}

\begin{figure*}
\includegraphics[scale=0.65,angle=90]{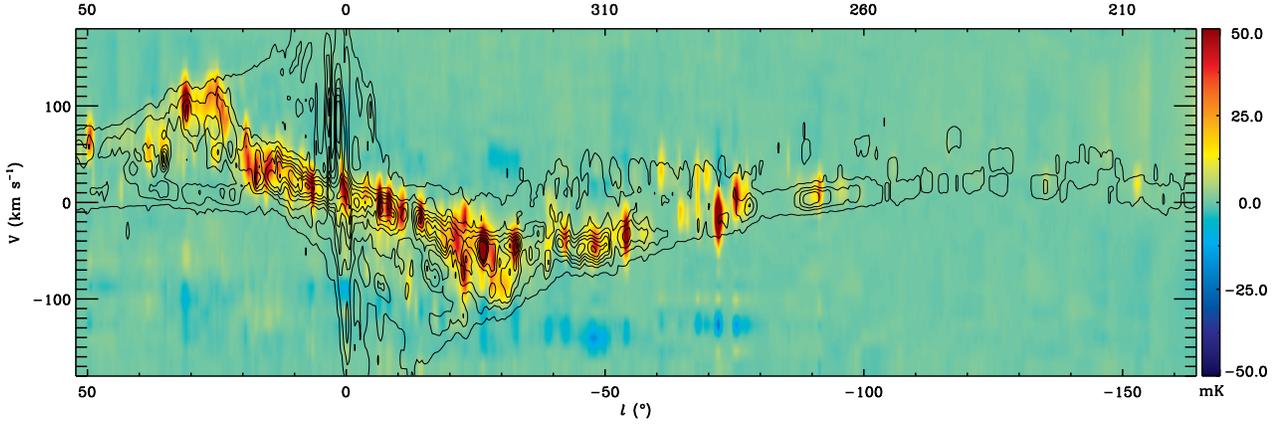}
\caption{Longitude-velocity diagram of the RRL brightness temperature
  integrated over a strip of $4\degr$ in latitude, centred in the
  Galactic plane. The spectra are averaged every
  $0\pdeg5$ in longitude. The contours show the distribution of the CO
  emission from the survey of \citet{Dame:2001}; they are at
    every 300\,mK from 25 to 1825\,mK. Both the RRL and the CO
  emission follow the velocity pattern
  of the spiral arms, with a wider velocity distribution for the
  molecular gas. \label{fig:lonvel}}
\end{figure*}

The RRL emission at eight velocity channels, between $V=-80.3$ and 107.1\kms\
is shown in Fig. \ref{fig:cmaps1}. Individual
\hii~regions can be identified in the different maps, as well as the
extended emission around and between them. A list of \hii~sources is presented in
Section \ref{sec:hiicat}. Beyond the inner Galactic plane, or at $|l|
\geq 60\degr$, there are only a few isolated complexes of \hii~regions
visible in the map, namely RCW57 G291.6-0.4, the Carina 
Nebula G287.4-0.6, RCW49 G284.3-0.3, and RCW38 G267.9-1.0, in the
Carina spiral arm, as well as the Rosette Nebula
G206.4-2.3 in the Perseus spiral arm. On the other
hand, emission at all velocities is visible across the inner Galaxy,
consistent with the fact that the bulk of star formation is occurring
in the inner two spiral arms of the Galaxy \citep{Wood:1989,Bronfman:2000}.
At velocities $V \ltsim 30$\kms\ the emission is broader in latitude since it arises from the
nearby Local and Sagittarius/Carina spiral arms. Emission further away from
us, in the Scutum/Norma spiral arms, has higher velocities and a narrower latitude distribution.
This is not the case in the
central $2\degr \times 2\degr$ of the Galaxy, where the RRL emission
is at velocities close to zero. However, this line-emitting gas is not necessarily
foreground or local gas but is also in the Galactic centre (GC) region \citep{Law:2009}.
This region will be further discussed in Section \ref{sec:gcl}.

\subsection{Longitude-velocity distribution}
\label{subsec:gasvel}

The distribution of the RRL velocity as a function of longitude is 
shown in Fig. \ref{fig:lonvel}, where we can see the change from positive to negative
velocities as we go from quadrants I to IV. The negative
(forbidden) velocities in quadrant I, for example associated with
the \hii~complex W43 at $l=30\degr$ and $V\sim -50$\kms, originate from helium and carbon
RRLs (Section \ref{sec:hecrrls}). The negatives seen in the diagram are associated with bandpass
distortions by strong continuum sources, as described in
Section \ref{sec:effects}, thus they appear at the
same longitude as the strong emission but at a different velocity.

The highest absolute values of velocity observed at $|l| \sim 30\degr$
corresponds to the lines of sight crossing the tangent of the Scutum-Norma spiral
arm, at a distance of about 4\,kpc (using a solar Galactocentric distance
of 8.5\,kpc, \citealt{Kerr:1986}). The tangent point of the
Sagittarius arm at $l\sim50\degr$ is not well covered by the RRL
survey, nevertheless there is clear emission associated with the W51
complex at $V\sim 60$\kms. There is also RRL emission associated with
the quadrant IV of the Carina spiral arm, at $l=300\degr$ and
$V\sim30$\kms.

The contours in Fig. \ref{fig:lonvel} trace the CO emission
detected in the survey of \citet{Dame:2001}. The CO spectra are
averaged within the same $0\pdeg5 (l) \times 4\degr (b)$ areas as the
RRL data, and smoothed from their initial resolution of 2\kms\ to
20\kms. The diagram clearly shows that the RRL emission is
mostly associated with the CO gas from the molecular ring, the region
of high CO emission within $|l| \ltsim 30\degr$. Furthermore, the fact
that star formation is not present in every molecular complex across
the Galaxy (e.g., \citealt{Evans:1991}) is well illustrated by the narrower velocity range covered by
the RRL emission, compared to the wider velocity spread of the CO
emission at a given longitude. There are peaks in RRL emission that do
not have an associated CO peak, e.g., at $l=284\degr$ and $V\sim0$\kms, $l=287\pdeg4$
and $V\sim -20$\kms, or $l=336\pdeg7$ and $V\sim-10$\kms. 
The second example corresponds to the Carina nebula, a giant
\hii~region with a large number of massive stars \citep{Smith:2008}. It is thus possible
that the parent molecular cloud has been disrupted by the powerful
stellar feedback from several generations of stars. 

The 3-kpc expanding arm \citep{Cohen:1976} is seen in CO \citep{Dame:2008}
emission within $|l| \ltsim 10\degr$ and at negative velocities, close
to the first contour in Fig. \ref{fig:lonvel}. 
High signal-to-noise RRL emission associated with the 3-kpc arm is
detected, but not seen in the longitude-diagram due to the large
(4\degr) latitude band used in averaging the spectra. 
Non-circular motions in the GC with a large spread of velocities, up to $\sim 200$\kms, have been
observed, both in molecular and atomic tracers \citep{Burton:1985a,Burton:1988,Dame:2001}.
However, as mentioned in the previous section, we find that the RRL
emission towards the GC is mostly at velocities
near zero, which is expected for circular motions nearly perpendicular
to the line of sight. Spectra towards the GC will be presented in
Section \ref{sec:gcl}, which also show the negative velocity component arising from the 3-kpc expanding arm.

\subsection{RRL integrated intensity}
\label{subsec:rrlint}

\begin{figure*}
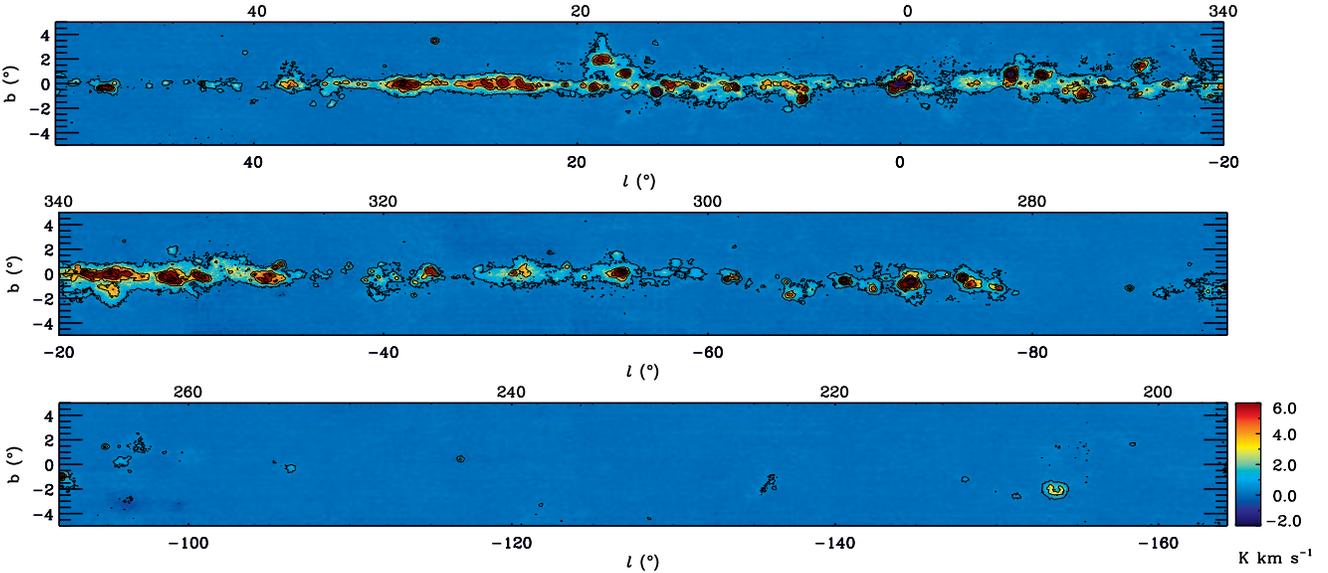

\hspace*{-1cm}\includegraphics[scale=0.92,angle=90]{Figs/fig7a.epsi}
\hspace*{-1.1cm}\includegraphics[scale=0.92,angle=90]{Figs/fig7b.epsi}
\includegraphics[scale=0.92,angle=90]{Figs/fig7c.epsi}
\caption{Map of the total RRL integrated emission in units of
  K\kms~and at a resolution of 14.4\,arcmin. The contours correspond to
  the free-free brightness temperature distribution at 1.4\,GHz
  (Section \ref{subsec:ff}); they are at 1.5, 7.5, 13.5, 25,
    45, 65, and 85\,K. The longitude coverage of the survey is divided into
   three sections as in Fig. \ref{fig:rmsmap}. \label{fig:intmap}}
\end{figure*}

The total RRL integrated emission for the Galactic plane region $l=196\degr$ --
$0\degr$ -- $52\degr$ and $|b| \leq 5\degr$ is shown in
Fig. \ref{fig:intmap}. This represents the first fully-sampled map of ionized
emission in this region of the Galaxy. The conversion of the line
integral into a free-free brightness temperature is presented in the
next section. 

The map of Fig. \ref{fig:intmap} is in units of K\kms\ and at a
resolution of 14.4\,arcmin. The velocity integration range is set
differently for spectra with peak temperature above and below
$3\sigma_{\rm T}$, where $\sigma_{\rm T}=4.5$\,mK is the typical rms noise measured in the
Galactic plane (Section \ref{sec:effects}).
This threshold roughly defines the ridge of 
emission within the first contour of Fig. \ref{fig:intmap}, where the 
spectra are integrated over 170\kms\ about their peak velocity. Such a
velocity range is wide enough to
include the emission allowed by Galactic rotation along the plane, without
including emission from He/C RRLs (Section \ref{sec:hecrrls}). 
In the case where the line is below $3\sigma_{\rm T}$ the central velocity is
assumed to be 0\kms, the velocity of local emission, and each spectrum
is summed within a narrower range of 92\kms. Finally, the data
  are clipped at $-3\sigma$, i.e., channels 
  with temperature below $-3$ times the spectral noise (Fig. \ref{fig:rmsmap}) are not
  included in the line integrated emission. We note that if we use a
  constant clip value of $-3\sigma_{\rm T}$, the resulting RRL integrated
  emission is about 3 per cent higher in the brightest regions of the
  inner Galaxy (within the first contour of
  Fig. \ref{fig:intmap}). 

The maximum and minimum of the map are $-4$ and 38\,K\kms, respectively, and the
rms noise measured away from the Galactic plane ($|b| \gtsim 3\degr$)
is $0.1$\,K\kms. The fraction of pixels with RRL integrated emission
below $-0.3$\,K\kms\ is only 0.4 per cent.
The dynamic range of the map, defined as the ratio
between the maximum intensity and the noise level, is 380.

\subsection{Free-free emission at 1.4\,GHz}
\label{subsec:ff}

The thermal continuum emission is obtained as described in Section
\ref{sec:introduction} (Eq. (\ref{eq:ff})), namely through the RRL integral and a value for the electron
temperature. The free-free emission is shown by the contours of
Fig. \ref{fig:intmap}, which follow the RRL integrated emission. 

The electron temperature of \hii~regions is known to vary
with the Galactocentric radius $R_{\rm G}$ \citep{Shaver:1983,Paladini:2004}: $T_{\rm e}$
decreases towards the Galactic centre due to its higher metal
content. In \paper\ we studied 15 \hii~regions within
$l=20\degr$ -- $44\degr$ and $|b| \leq 4\degr$ and found that:
\begin{equation}
T_{\rm e}[\rm K] = (3609 \pm 479) + (496 \pm 100)R_{\rm G}[\rm kpc]
\label{eq:te}
\end{equation}
using the \citet{Fich:1989} rotation curve, with
$\Theta_{\odot}=220$\kms~and $R_{\odot}=8.5$~kpc, the rotation
velocity and Galactocentric distance of the Sun, respectively\footnote{With the adopted rotation
    curve, the Galactocentric radius is given by} $R_{\rm G} = \frac{221.64R_{\odot}}{\Theta_{\odot} +
  V/{\rm sin}(\ell) + 0.44R_{\odot}}$.
\citep{Kerr:1986}. This results in electron
temperatures about 5 -- 10 per cent higher than those
from both \citet{Shaver:1983} and
\citet{Paladini:2004}. Furthermore, we have shown in \paper\  that for that
same region of the Galaxy, the electron temperature of the diffuse free-free emission is
similar to that of the individual \hii~regions. 
Therefore, we apply
Eq. (\ref{eq:te}) to every pixel in the RRL
spectral cube where the peak temperature is above $3\sigma_{\rm T}$, as defined in
Section \ref{subsec:rrlint}, using the central line
velocity to estimate $R_{\rm G}$. Note that for each spectrum
  the central velocity corresponds to the brightest peak; when multiple
  peaks are present, meaning when there is more than one velocity
  component along the line of sight, $T_{\rm e}$ should be estimated
  based on the relative brightness of each of the peaks. Nevertheless,
  neglecting the contribution of different velocity components does not significantly affect the
derived value of $T_{\rm e}$ within the typical uncertainty of 1000\,K,
associated with Eq.~(\ref{eq:te}).
The typical value of electron
temperature obtained along the Galactic plane is 6000\,K.
For the low signal-to-noise pixels,
for which we ascribed the local velocity of 0\kms, we set $T_{\rm
  e}=7000$\,K, the mean electron temperature in the Solar
neighbourhood
(\citealt{Shaver:1983,Paladini:2004}, \paper). The exact value of
$T_{\rm e}$ in these pixels is not crucial since their emission is
within the noise level. Due to the complexity of the GC
region and the difficulty of deriving the Galactocentric distance based
on the radial velocity at these longitudes, we do not apply Eq. (\ref{eq:te}) within
$l=354\degr$ -- $0\degr$ -- $4\degr$ but set $T_{\rm e}$ to a constant
value of $T_{\rm e}=6000$\,K to obtain a smooth variation of the 
electron temperature with longitude. However, we note that lower
values, $T_{\rm e} \ltsim 4000$\,K, have been determined towards the
GC \citep{Law:2009}, which are consistent with the result
of Eq.~(\ref{eq:te}) for $R_{\rm G}=0$.

For typical values of $T_{\rm e} = 6000 \pm 1000$\,K, the uncertainty in the derived
free-free continuum emission is 20 per cent. If a constant
electron temperature of 7000\,K is used throughout the whole Galactic
plane, then the free-free map is simply a scaled version of the RRL
integrated emission, with 1.0\,K\kms\ equivalent to 2.8\,K of
brightness temperature at 1.4\,GHz (Eq. (\ref{eq:ff})). The minimum and
maximum of the free-free map are $-9$ and 116\,K, respectively, and the rms
noise measured away from the Galactic plane is 0.3\,K. Less than
0.3 per cent of the pixels have emission below $-0.9$\,K. 

\begin{figure*}
\centering
\includegraphics[scale=0.45]{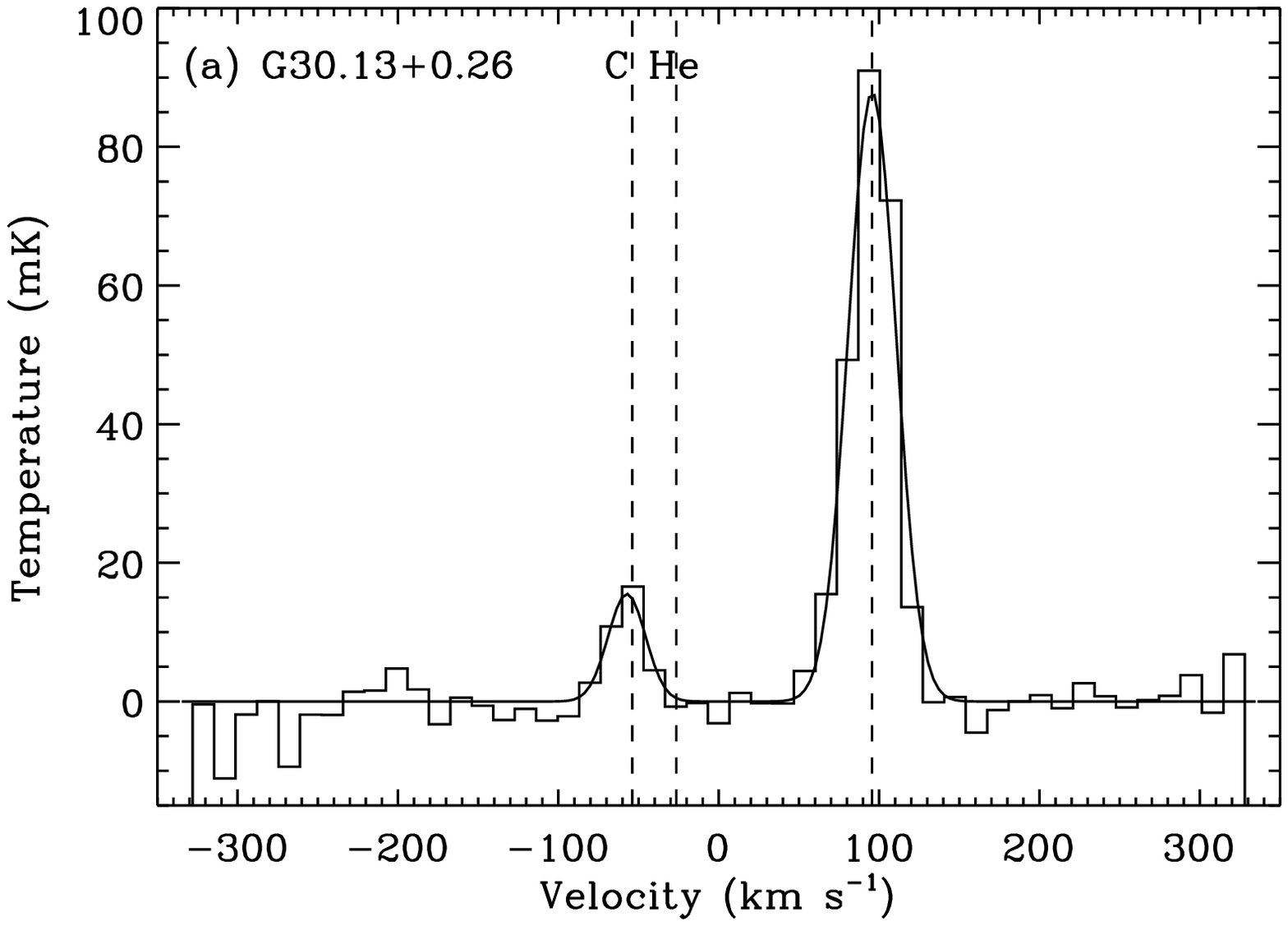}
\includegraphics[scale=0.45]{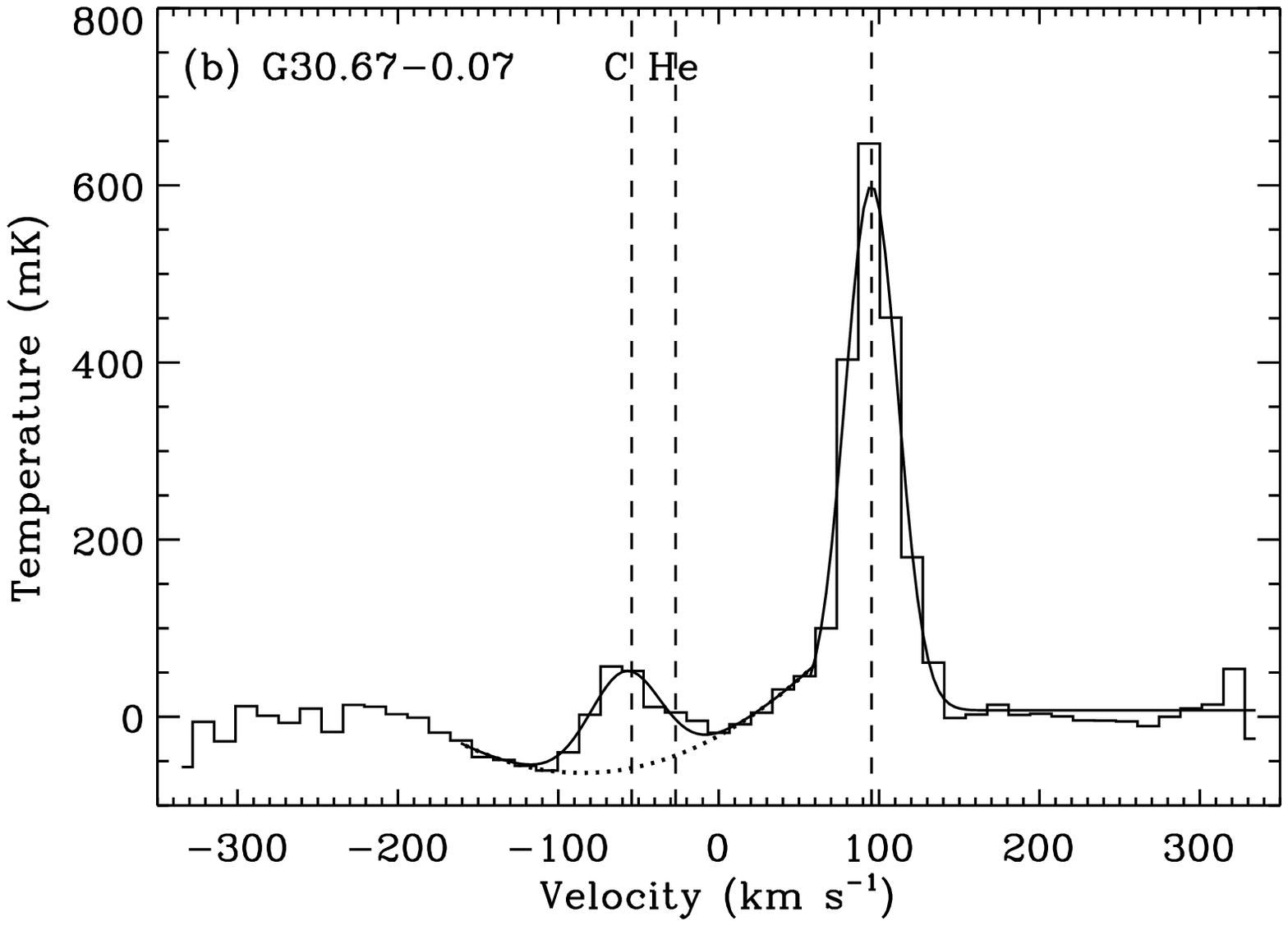}
\caption{Average spectra, $12\times12$\,arcmin$^{2}$, measured in a
  diffuse line of sight (a), next to the bright \hii~region W34
  (b). The dashed vertical lines indicate the expected velocities of
  the He and C RRLs, relative to the H line. The two component
  Gaussian fit to each spectrum is given by the solid line. In panel (b), the dotted
  curve corresponds to the second order fit performed to correct the baseline. \label{fig:hecspec}}
\end{figure*}
\begin{figure*}
\centering
\includegraphics[scale=0.9]{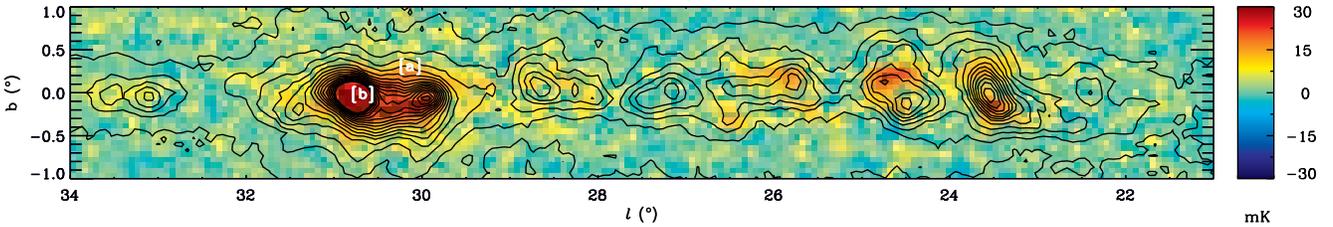}
\caption{Map of the RRL emission at $V=-55$\kms, the expected
  velocity of C line emission from the Scutum spiral arm. The
  contours correspond to the H RRL at $V=94$\kms\ and are at every
  25\,mK from 10 to 410\,mK. The two labels in the map show the
  positions at which the spectra of Fig. \ref{fig:hecspec} are measured. \label{fig:hecmaps}}
\end{figure*}

\subsection{Helium and Carbon RRLs}
\label{sec:hecrrls}

The 168$\alpha$ -- 166$\alpha$ RRLs from helium and carbon are also detected in
this survey. They are separated from the H line by $-122.1$ and 
$-149.5$\kms\ respectively, relative to the H RRL velocity scale. As
the velocity separation between the He and C RRLs is similar to the spectral
resolution of 20\kms, the two lines are usually blended in these data. The C RRL arises from
colder gas lying in front of the continuum source, in photodissociation regions
\citep{Balick:1974,Dupree:1974}. Stimulated emission by the
background source results in line temperatures that are orders of magnitude
above that expected from the carbon abundance \citep{Goldberg:1967}. If the cold gas is not associated with the
\hii~region that induces the line enhancement but is at a different
velocity, then the C RRL is correspondingly shifted in velocity and
can be separated from the He line (\paper). C RRL emission can also
arise from diffuse gas \citep{Kantharia:2001}, ionized by
the interstellar UV radiation field. 

The spectra of Fig.~\ref{fig:hecspec} show the He/C RRLs measured
towards the W43 G30.7+0.0 complex. Fig.~\ref{fig:hecspec} (a)
is the spectrum for a diffuse line of sight, $0\pdeg6$ away from the centre
of the bright \hii~region. Gaussian fits
give a width of $29\pm2$\kms\ for the H line and $19\pm6$\kms\ for the RRL
at the velocity expected for the C line; the narrower width of the latter is
consistent with a line arising from a cooler medium. The ratio between
the C and H RRL amplitudes is $17\pm2$ per cent. An example of the
distortion of the bandpass (Section~\ref{sec:datanalaysis}) by the strong continuum
source W43 is provided by the spectrum
of Fig.~\ref{fig:hecspec} (b). After a second order polynomial fit
around the carbon central velocity, we find a large line width of
$48\pm5$\kms, indicating the presence of both He and C RRLs. 
Their combined amplitude of about 110\,mK results in a ratio of 18 per
cent relative to the H line. The ionization energy of atomic
carbon is 11.3\,eV, lower than that of H, 13.6\,eV, and much lower
than that of He, 24.6\,eV. Our observations are consistent with the
fact that He RRLs should be detected towards the \hii~region, where the stellar radiation is
sufficiently high, whereas C RRLs can also be observed further away from
the ionizing sources.

The channel map of Fig.~\ref{fig:hecmaps} shows the line emission at
$V=-55$\kms\ for the $l$-range $21\degr$ -- $34\degr$ and within 1\degr\
of the Galactic plane, where the H RRL peaks at about 94\kms.
Emission at the C RRL velocity
is seen across the whole W43 complex as well as towards the W41 (G23.4+0.0) and W42
(G25.3-0.1) \hii~regions, where the C line is brighter outside the
continuum peaks. 
Diffuse C{\sc ii} gas is best detected at low
frequencies, $\ltsim 1$\,GHz, where C RRLs from the Galactic plane
have been observed mostly within $l<20\degr$, both in emission and
absorption \citep{Erickson:1995,Kantharia:2001}. The decrease in the
background continuum, thus in the stimulated emission, 
along with the low angular resolution
of past surveys, are responsible for the limited number of detections of C
RRLs beyond $l=20\degr$ \citep{Roshi:2000, Roshi:2002}.

Here we present the first
observations of C RRLs at 1.4\,GHz from diffuse gas along the Galactic
plane. The Galactic plane region shown in Fig.~\ref{fig:hecmaps} is one of
the cleanest in which to identify the C RRL as it shows only one H RRL component, consistent
with the velocity of the Scutum spiral arm. In the fourth quadrant of
the Galaxy, specifically within $l=270\degr$ -- 330\degr, negative spikes at
$\sim -130$\kms\ corrupt the spectra at or near the expected velocity
of the He/C RRLs (Section~\ref{sec:effects}). However, these lines are
observed towards other objects across the whole coverage of this
survey and can be investigated.

Studies of the morphology and physical conditions of C{\sc ii} regions
can be pursued using the present data, combined with previous surveys,
not only of C RRLs at different frequencies but also of other cold gas
tracers such as \hi\ and C$^{+}$.

\section{A catalogue of \hii~regions}
\label{sec:hiicat}

In \paper\ we presented a list of 48 \hii~sources in the region
$l=20\degr$ -- $44\degr$, $|b| \leq 4\degr$, with sizes, flux
densities, and velocities as measured from the RRL free-free map. All those objects are
known and have been observed by many continuum as well as RRL surveys
(e.g., the source compilation by \citealt{Paladini:2003} and the
recent RRL survey by \citealt{Anderson:2011}). 
Here we extend this list to include all the thermal sources detected
in the present free-free map. This list gives the
source parameters, free of any synchrotron contamination, which can
occur when analysing total continuum surveys, and due to its
resolution of 
$\sim14$\,arcmin, it includes the more extended emission around and between
features. An additional advantage of RRL detections is that for each source we have a
velocity measurement, thus its distance to the Sun can be obtained for diffuse and compact \hii\ regions.

\subsection{Source detection}
\label{sec:catdetection}

We follow the same approach as in \paper\ to create a catalogue of
\hii~regions from this survey, using the source extractor
algorithm SE{\sc xtractor} \citep{Bertin:1996} and adjusting its main input parameters
such as background mesh, detection threshold and filter
function. These are tuned to recover not only the bright sources easily visible in the
map, but also weaker objects, whose spectra are also checked to confirm
whether they do indeed correspond to a real signal. The background is estimated
in every $1\degr \times 1\degr$ region, utilising an
iterative clipping of the background histogram until convergence at $\pm3\sigma$
around its median. We then apply a  $3\times3$\,deg$^{2}$ median filter
to create a smoother background map. We note that the typical
size of the detected sources is $\sim0\pdeg1$, thus significantly smaller than the
background mesh size. The source detection is performed on the input
free-free map after filtering with a
Mexican Hat filter of width 12\,arcmin, similar to the beam size. If a
source extends beyond 3 pixels, or $3\times4$\,arcmin, and is above
the detection threshold of 2$\sigma$, where $\sigma = 0.3$\,K
(Section~\ref{subsec:ff}), it is considered a detection. The source
size is measured from the second-order moment map and given in 
terms of $a$ and $b$, the ellipse major and minor
axes. We compute the flux
density $S$ of each source assuming a Gaussian profile as follows
\begin{equation}
S = S_{\rm p} \frac{a \times b}{fwhm^{2}}
\label{eq:fluxdens}
\end{equation} 
where $S_{\rm p}$ is the peak flux and $fwhm$ is the beam FWHM of
14.4\,arcmin. The peak flux is measured at the central position of
each object from the background-subtracted free-free map.

\subsection{Properties of the catalogue}
\label{sec:catprop}

\begin{figure*}
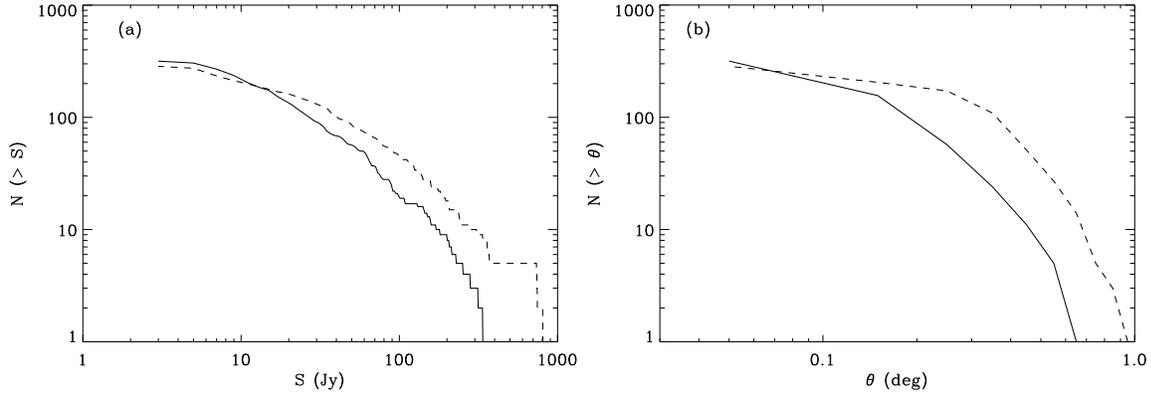

\centering
\includegraphics[scale=0.45]{Figs/fig10a.epsi}
\includegraphics[scale=0.45]{Figs/fig10b.epsi}
\caption{Cumulative counts as a function of the (a) flux density and
  (b) size. The solid line corresponds to the 317 sources listed in Table
  \ref{tab:hiicat} and the dashed line to the 285 objects detected
  when using a Mexican Hat filter of 20\,arcmin width.\label{fig:comp}}
\end{figure*}

The first 10 rows of the catalogue of sources extracted from
  the free-free map are reported in Table~\ref{tab:hiicat}. This table
  lists the Galactic coordinates, angular size, position angle, peak
  and integrated flux density, and central velocity of each object.
The broadening of the beam is difficult to measure for
small sources, especially if they are also weak. We consider that a
source is unresolved when its size is smaller than 10\,arcmin, which
implies an angular broadening of 3\,arcmin relative to the 14.4\,arcmin
beam. We take this conservative value, which is much larger than the variation of
the FWHM of the individual beams (0.5\,arcmin, Table \ref{table1}), to
account also for beam modelling uncertainties: the typical uncertainty
on the fitted sizes is 2\,arcmin, which follows from both the noise
level in the free-free map, as well as the uncertainty on the background
estimation. Therefore, unresolved sources are indicated with the symbol $p$ whenever the observed
size is below 17.5\,arcmin. Further, if the area of a source
($a\times b$) is smaller than the beam area ($fwhm^{2}$), $S=S_{\rm
  p}$. The position angle (PA) is undefined in such cases. The
uncertainty on the peak flux takes into account the spectral noise of the RRL data as well as the
rms of the background map estimated by SE{\sc xtractor}. We do not
include the systematic uncertainty associated with the electron
temperature nor the overall calibration uncertainty of the survey. The
uncertainty on the flux densities is about 25 per cent.

Table~\ref{tab:hiicat} includes sources that have a
signal-to-noise ratio on the flux density above 2. Fig.~\ref{fig:comp} (a) shows the integral
count $N(>S)$ for the 317 objects as a function of the flux density
$S$. The catalogue has a lower limit of 3\,Jy and is complete down to
5\,Jy at a 95 per cent level. This is partly due to source
confusion, meaning the difficulty of resolving the Galactic plane
structure into individual objects, as well as to the sensitivity of
the survey. The cumulative distribution of the
size $\theta$ of the \hii~regions is shown in Fig.~\ref{fig:comp}
(b), where $\theta = (\theta_{a}+\theta_{b})/2$ and $\theta_{a}$ and
$\theta_{b}$ are the measured values $a$ and $b$ deconvolved with the beam. We compare these results
with those obtained when using a Mexican Hat filter of larger width,
20\,arcmin. Fig.~\ref{fig:comp} shows that both the completeness and lower
limit of the flux densities and sizes are not affected, meaning that
using a filter function of width larger than the beam FWHM does not
necessarily recover fainter sources. On the other
hand, the larger convolving beam results in larger sizes for the 285
detected objects. As a consequence, the flux densities are also
higher. The total flux of the 317
\hii~regions in Table~\ref{tab:hiicat} is 121\,212\,Jy\,beam$^{-1}$, 73 per cent of which arises from the
inner Galaxy, $|l| \le 50\degr$. The contribution of the individual
sources to the total free-free emission (Fig. \ref{fig:intmap}), which amounts to 415\,371
\,Jy\,beam$^{-1}$, is about 30 per cent. This result is similar to
that found in \paper\, for the smaller longitude range
analysed. We note that some of the sources in
the list are part of large \hii~complexes (e.g., W42, W47) or
are usually identified as a single object (e.g., sources 110
  -- 114 correspond to the Rosette Nebula). Moreover, due to the
  presence of saturated pixels at their centres, the 6 \hii~regions
  marked in Fig. \ref{fig:sensmap} are not recovered from the free-free map, and
  thus not listed in Table \ref{tab:hiicat}.

The last column of Table~\ref{tab:hiicat} gives general remarks on the
object as well as its commonly used name(s). These notes are taken
from the \citet{Paladini:2003} Master Catalogue. Since the \citeauthor{Paladini:2003} catalogue is a
compilation of results from several surveys with beam sizes generally
smaller than the HIPASS 14.4\,arcmin FWHM, we consider a match if the distance between objects
from both catalogues is within 7.5\,arcmin, about half of the beam
FWHM. A total of 195 objects has one or more spatially
  corresponding source in the Master
  Catalogue. The central RRL velocity listed in Table~\ref{tab:hiicat}
is derived from a one- or two-component Gaussian fit to the average $12\times
12$\,arcmin$^{2}$ spectrum measured at the position of each object.
There are 374 discrete H RRL components from the 317 sources in the
list. We also check our estimates of the RRL velocity against those in
the Master Catalogue, when available. The agreement is better than
6.7\kms, or half of a channel, for 84 per cent of the objects with
counterparts in the \citeauthor{Paladini:2003} list.
Due to the difference in angular resolution between these surveys, the
flux densities also differ: the values found in our
work can be a factor of 3 -- 5 times higher than those given in the
Master Catalogue, after scaling from 2.7 to 1.4\,GHz with a flux
spectral index $\alpha=-0.1$ ($S\propto \nu^{\alpha}$,
\citealt{D3:2003}, \citealt{Draine:2011}).

The present catalogue lists 50 objects within the region
$l=20\degr$ -- $44\degr$, $|b| \leq 4\degr$, compared to the 48 included
in the \hii~region list of \paper. This difference involves objects
that have a peak flux below 15\,Jy\,beam$^{-1}$, thus below 3 times
the completeness limit of the survey. The flux densities in both
catalogues agree within the errors; the mean value of the ratio
between the flux densities from this work and from \paper\ is 1.3,
when considering the 16 sources that have a peak flux above
15\,Jy\,beam$^{-1}$. The discrepancy follows from a combination of
factors: (i) data reduction, especially concerning the iterative gridding
used here, which mostly affects the flux density of the smaller
sources; (ii) data analysis, we have followed a different method to
integrate the lines and produce the free-free map; (iii) the
calibration has also changed relative to \paper; (iv) some of the sizes fitted
by SE{\sc xtractor} are somewhat higher, which increases the flux
densities. 

\begin{figure*}
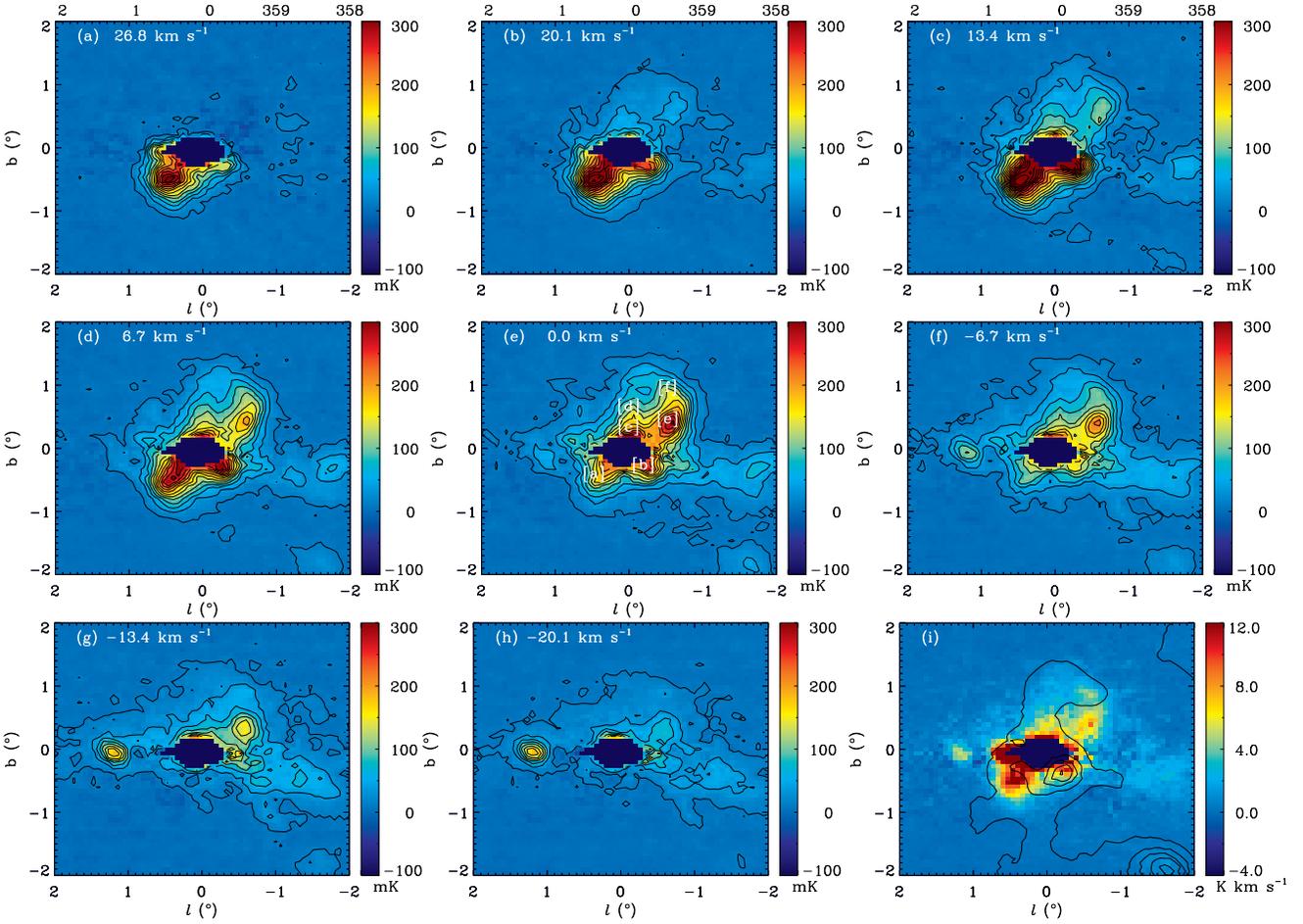

\includegraphics[scale=0.45]{Figs/fig11a.epsi}
\includegraphics[scale=0.45]{Figs/fig11b.epsi}
\includegraphics[scale=0.45]{Figs/fig11c.epsi}
\includegraphics[scale=0.45]{Figs/fig11d.epsi}
\includegraphics[scale=0.45]{Figs/fig11e.epsi}
\includegraphics[scale=0.45]{Figs/fig11f.epsi}
\hspace*{0.15cm}\includegraphics[scale=0.45]{Figs/fig11g.epsi}
\includegraphics[scale=0.45]{Figs/fig11h.epsi}
\includegraphics[scale=0.45]{Figs/fig11i.epsi}
\caption{Channel maps of the RRL emission towards the GC region at several
  velocities, from panels (a) to (h). The southern part
  of the lobe is seen to peak at positive velocities, whereas the
  northern part is brightest at $V \sim 0$\kms. The
  contours are at every 25\,mK from 10 to 410\,mK. We note that
  the central pixels in these maps are blanked due to the
  saturation of the receiver electronics by the strong GC continuum emission
  (Section~\ref{sec:effects}). The labels in panel (e) are at the
  positions where the spectra of Fig.~\ref{fig:gclspec} are
  taken. Panel (i) shows the RRL integrated emission map of the GC
  region with contours from the velocity-integrated \ha~emission;
  these are at every 50\,R from 25 to 225\,R. \label{fig:gcl}}
\end{figure*}

\section{The Galactic centre}
\label{sec:gcl}

In this section we present the RRL data towards the GC
region. These are the first fully-sampled observations of the ionized
gas in RRL emission, which reveal a southern
counterpart to the northern GC lobe previously detected at radio frequencies
(Section~\ref{sec:gcsouth}). We also present evidence of diffuse
ionized gas in the expanding 3-kpc arm (Section \ref{sec:gc3kpc}).  

\begin{figure*}
\centering
\includegraphics[scale=0.32]{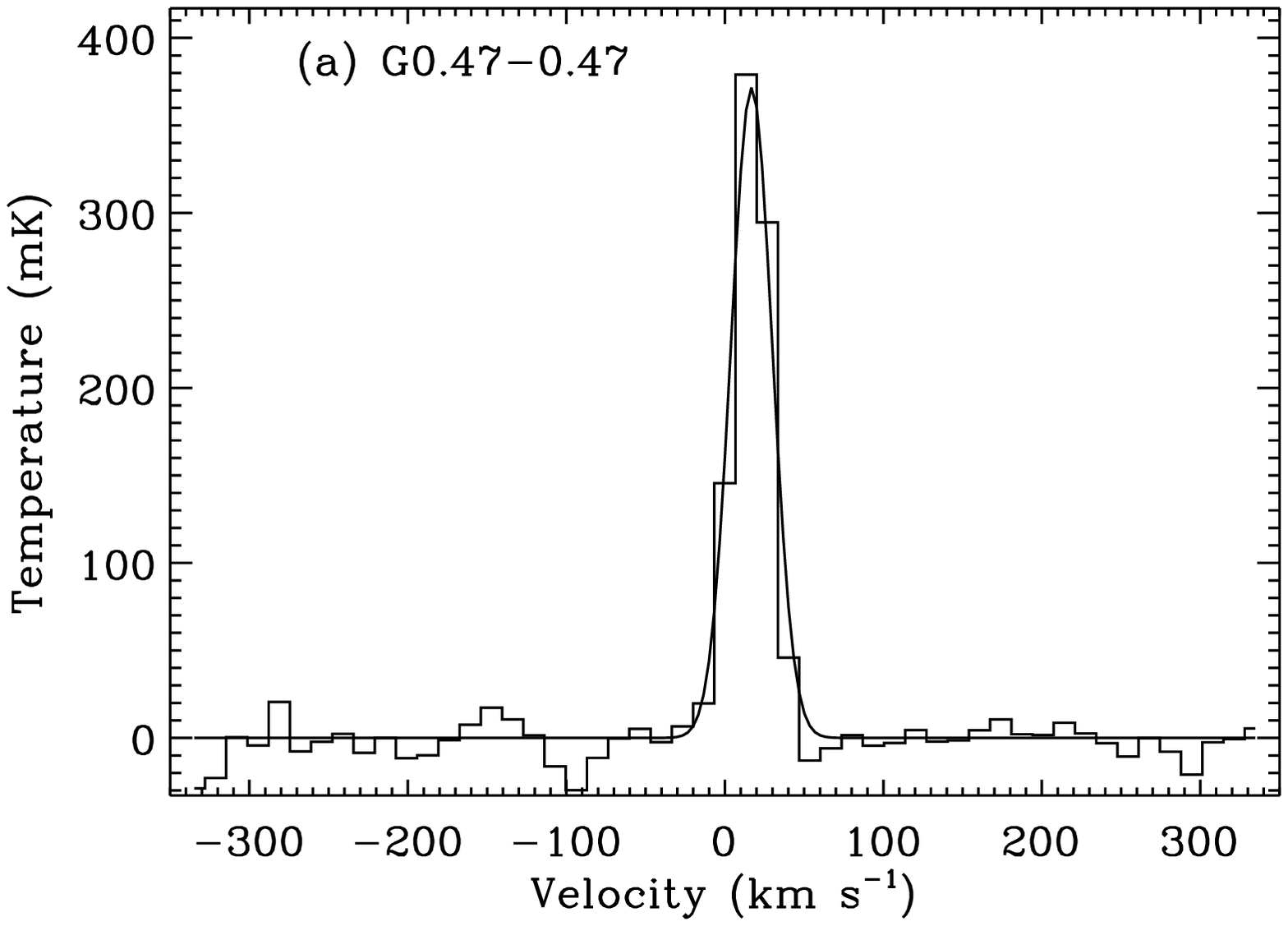}
\includegraphics[scale=0.32]{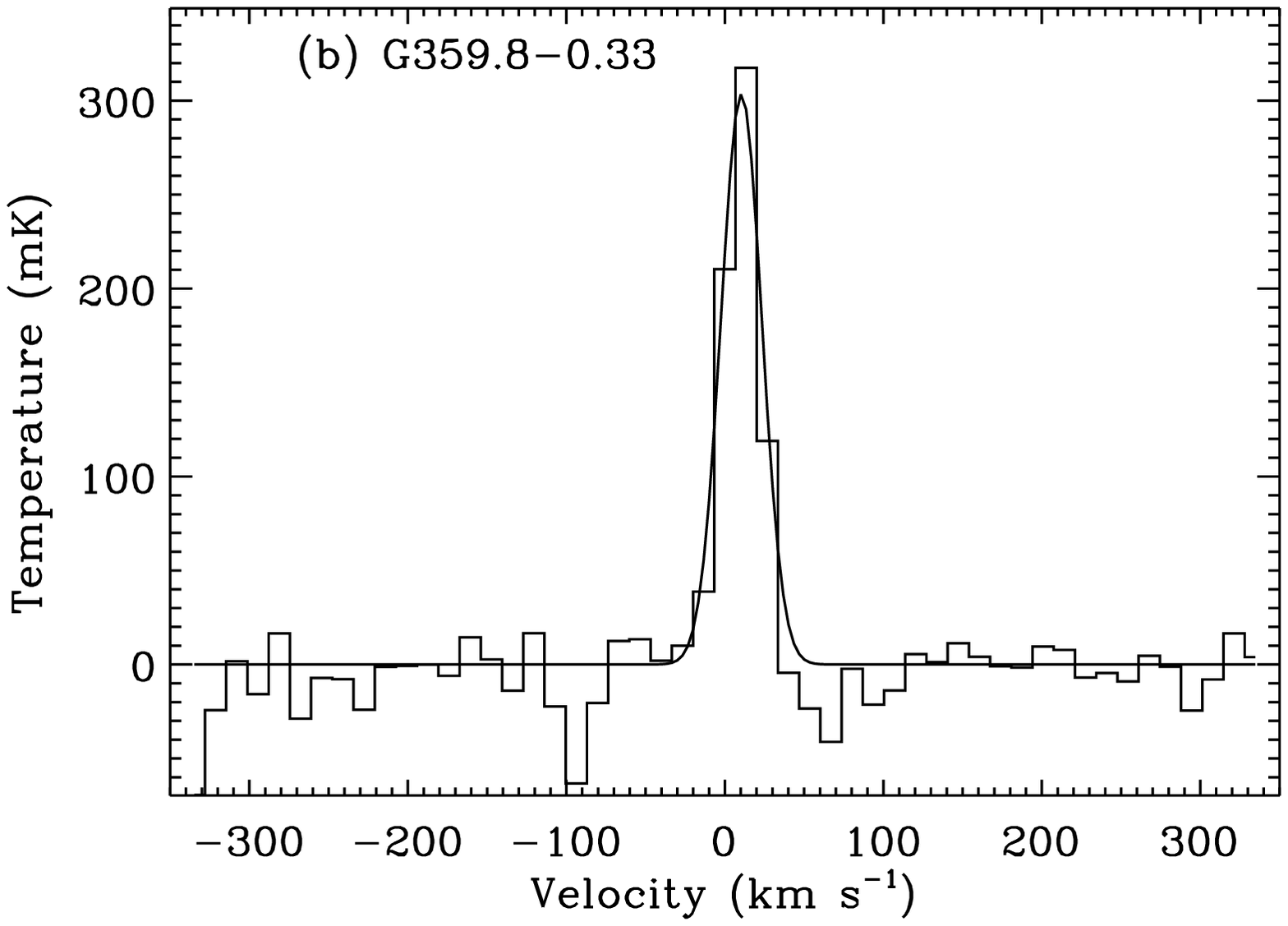}
\includegraphics[scale=0.32]{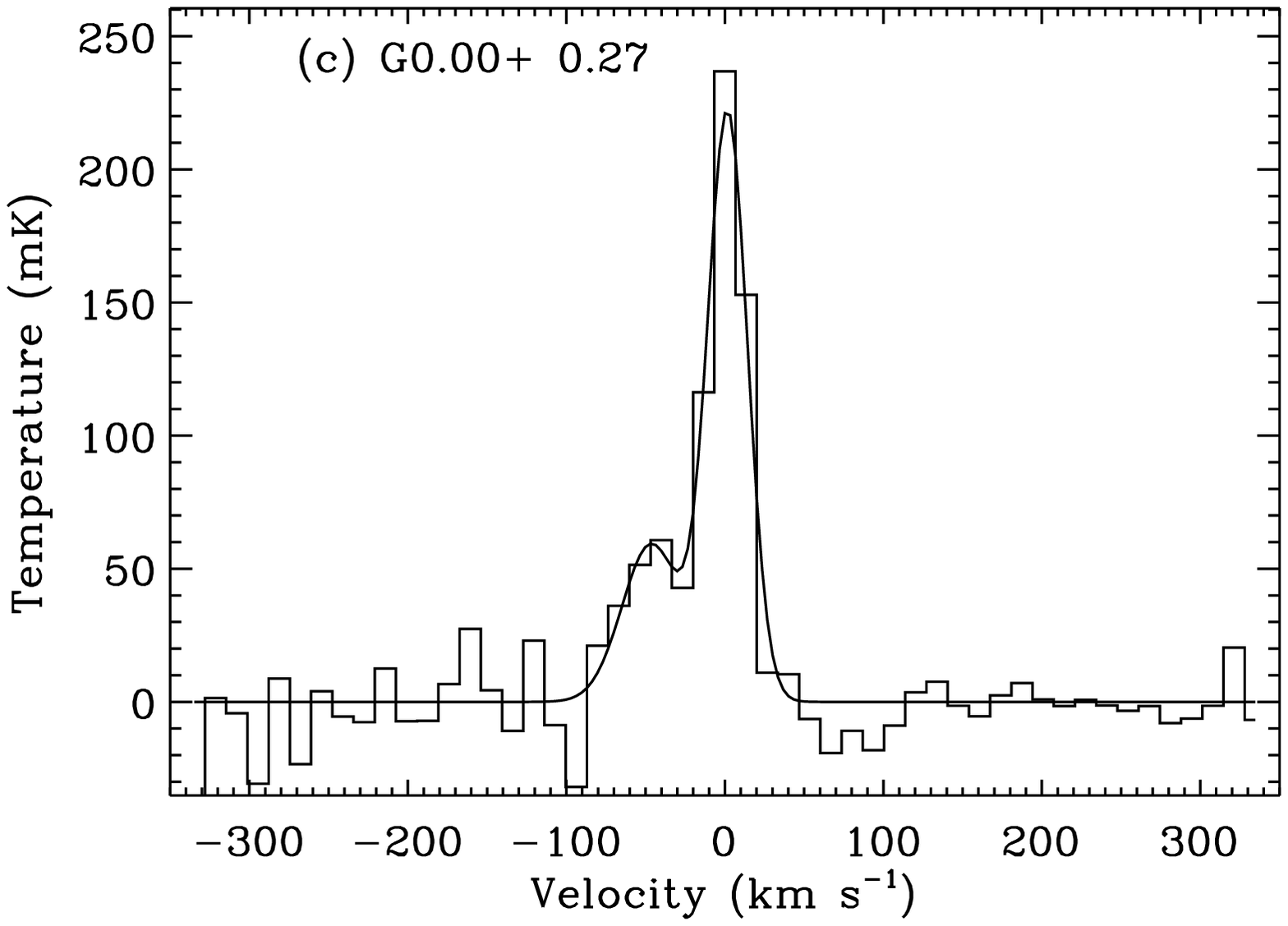}
\includegraphics[scale=0.32]{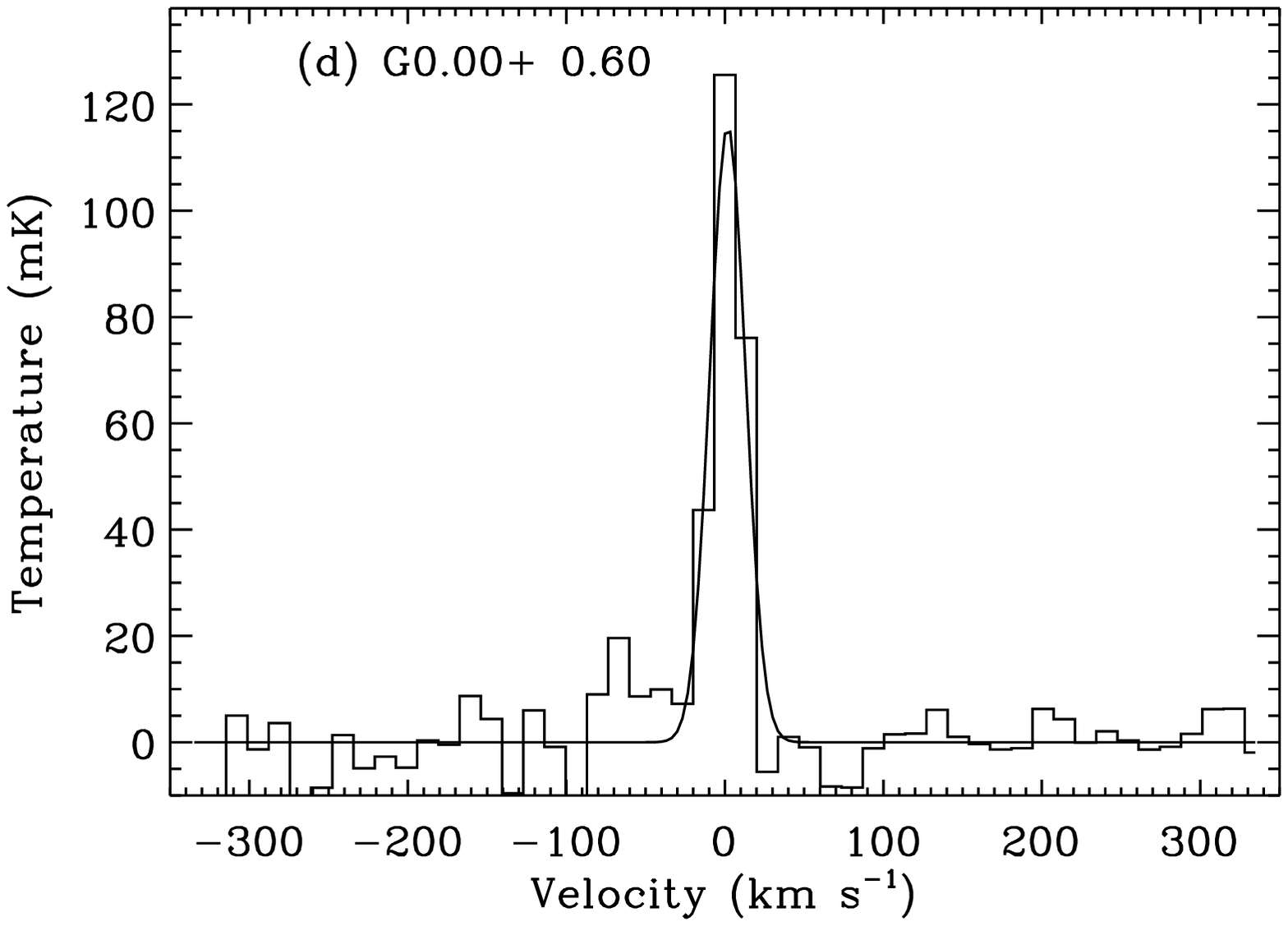}
\includegraphics[scale=0.32]{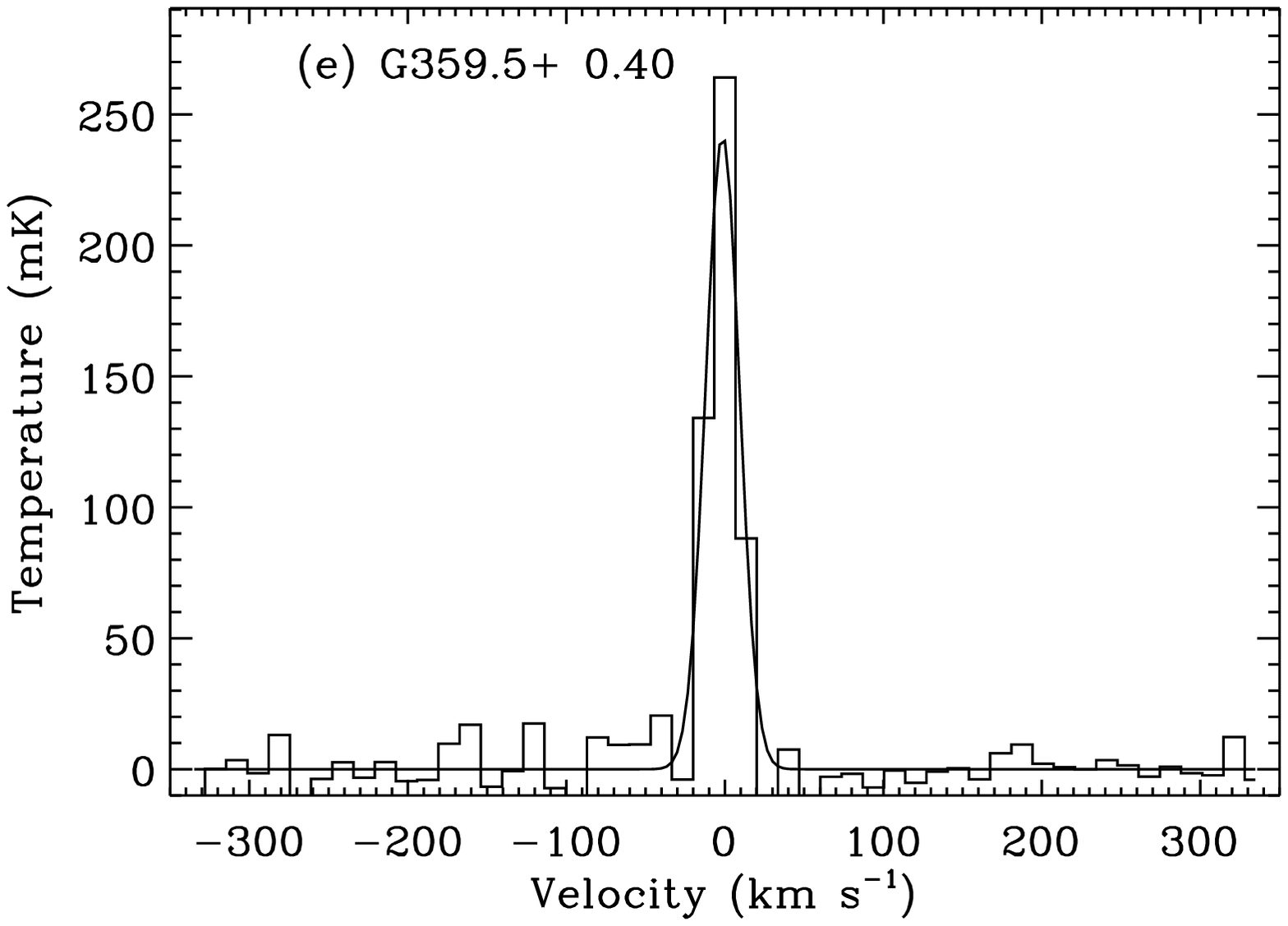}
\includegraphics[scale=0.32]{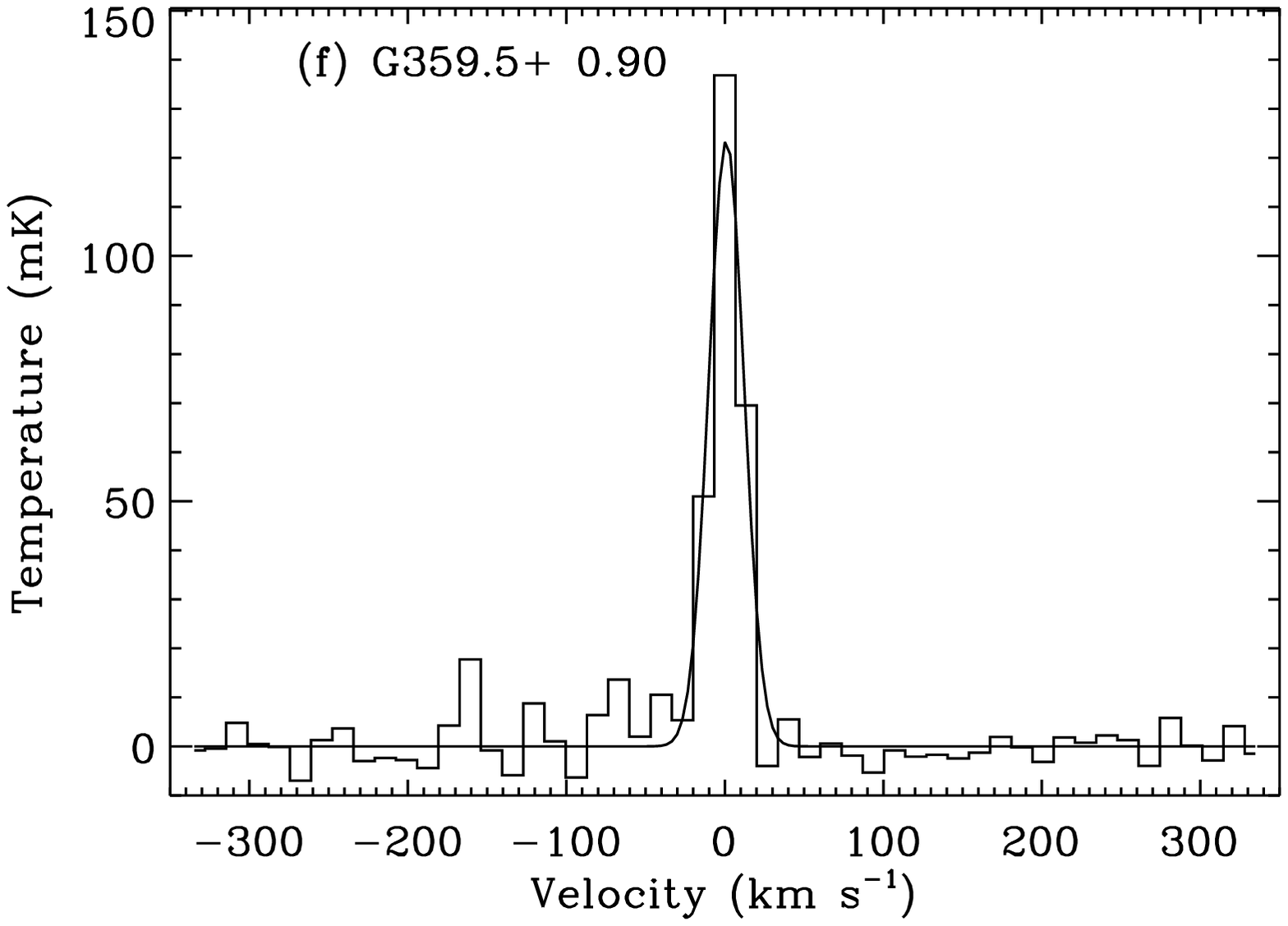}
\caption{Average spectra, $12 \times 12$\,arcmin$^{2}$, taken at
  positions in the GC lobe, as indicated in Fig.~\ref{fig:gcl} (e). 
Panels (a) and (b) correspond to the southern part
of the lobe, (c) and (d) to the east northern ridge,
and (e) and (f) to the west northern ridge. Gaussian
  fits to the spectra are shown. \label{fig:gclspec}}
\end{figure*}

\subsection{The southern lobe}
\label{sec:gcsouth}

The degree-scale lobe seen towards the GC has been studied by
\citet{Law:2009} using several RRL observations between 4.56
and 5.37\,GHz. This structure was first noted in a continuum survey at
10\,GHz by \citet{Sofue:1984}, who suggested that the lobe was of
thermal origin, possibly the result of an energetic outflow
phenomenon. The detection of RRL emission by \citet{Law:2009} confirmed
the thermal nature of the lobe, which is described as two vertical
edges near $(l,b)=(0\pdeg0,0\pdeg5)$ and
$(l,b)=(359\pdeg4,0\pdeg5)$, connected by a cap at $b\sim1\degr$, thus
north of the Galactic plane. These pointed observations
with the Green Bank Telescope (GBT) focused on the ridges as well as on
an horizontal strip from $l=359\pdeg25 \rightarrow 0\pdeg25$ at $b=0\pdeg45$ and
provided important information on the temperature and density of the
ionized gas in this structure: the line-emitting gas has low velocity,
$|V| \ltsim 10$\kms, and narrow widths, $\Delta V \sim 9-14$\kms. 
Here we provide the first fully sampled map of the ionized GC lobe,
which shows not only the known northern structure but also a
southern counterpart detected for the first time at radio frequencies.
\begin{figure}
\centering
\includegraphics[scale=0.7]{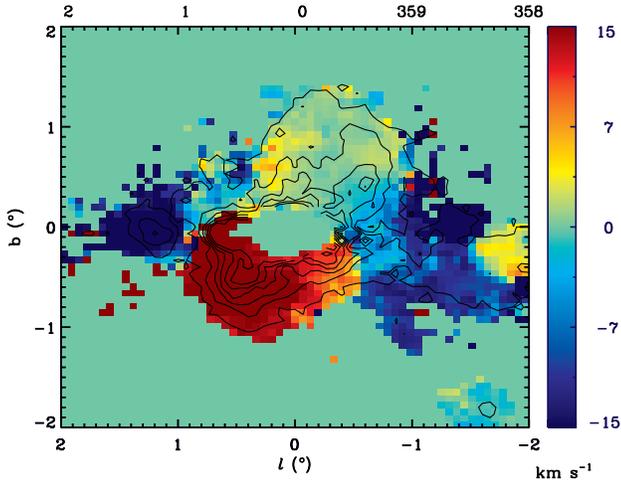}
\caption{Map of the velocity at peak of the RRL emission towards the
  inner $4\degr \times 4\degr$ of the Galaxy. The GC southern
lobe is seen at velocities of $\sim 15$\kms, whereas the northern part
is at $V \sim 0$\kms. The contours correspond to the RRL integrated
emission (Fig. \ref{fig:gcl} (i)) and are at every 2\,K\kms\ from
  1 to 11\,K\kms. \label{fig:gclvelmap}}
\end{figure}

Channel maps at some relevant velocities are
shown in Fig. \ref{fig:gcl}. The two ridges and cap are brightest in
Fig.~\ref{fig:gcl} (e), at $V \sim 0$\kms, whereas the southern
extension of the lobe is brighter in Fig.~\ref{fig:gcl} (c), at $V
\sim 13$\kms. Therefore, besides the 
small velocity gradient of  $\sim 10$\kms\ measured by \citet{Law:2009}
in the northern lobe, there is also a difference with relation to
the southern counterpart. This will be further discussed below (Fig. \ref{fig:gclspec}).
In Fig. \ref{fig:gcl} (g) the north-west ridge of
the lobe is still visible, as well as the \hii~region
G1.13-0.07. Fig. \ref{fig:gcl} (i) shows the contours of the
\ha~integrated emission towards the GC region overlaid on that of the
RRLs. The \ha~data are from the all-sky map composed by
\citet{Finkbeiner:2003}, which is based on the Southern
H-Alpha Sky Survey Atlas (SHASSA) survey
\citep{Gaustad:2001} in this region of the sky; these data are
smoothed from their initial angular resolution of 6\,arcmin to 14.4\,arcmin.
The fact that the optical line does not follow the
thermal emission from the lobes, indicates that these structures are
indeed at a large distance from us and hence are absorbed by the dust
along the long line-of-sight towards the GC. The \ha~emission peaks at
around $(l,b)=(359\pdeg75,-0\pdeg35)$, presumably a local \hii~region,
which is brightest in the RRL channel map of Fig.~\ref{fig:gcl}
(c). The second \ha~peak at $(l,b)=(358\pdeg3,-1\pdeg8)$ corresponds
to the RRL emission seen at $V=0$\kms\ in Fig.~\ref{fig:gcl} (e).

Spectra taken at positions across the GC lobe are shown in
Fig.~\ref{fig:gclspec}. These are $12 \times 12$\,arcmin$^{2}$ average
spectra, to which a single or double component Gaussian is fitted.
Figs. \ref{fig:gclspec} (a) and (b) show that the southern
ionized lobe is at positive and higher velocities, 17\kms\ and
10\kms\ respectively, than the north east ridge in
Figs.~\ref{fig:gclspec} (c) and (d), which has
a mean velocity of 2\kms. The spectra of
Figs.~\ref{fig:gclspec} (e) and (f), which
correspond to the west bright ridge, have central velocities of
$-2$ and 1\kms, respectively. These values have a typical uncertainty
of 1\kms, when taking the spectral noise into account.
All spectra
have narrow widths of about $19\pm4$\kms\ and are thus unresolved; the
uncertainty corresponds to the
rms scatter in the widths. These results are in
agreement with the higher resolution observations by \citet{Law:2009},
who detected two velocity components, positive and negative, in both
bright ridges.  We note, however, that owing to the higher
resolution of the GBT ($\sim 3$\,arcmin) the pointed observations of
\citet{Law:2009} towards each of the ridges lie approximately within
one beam of the present survey. Moreover, our velocity resolution of
20\kms~is also lower than that of GBT RRL survey, $1.5$\kms. The
second velocity component in the spectrum of Fig.~\ref{fig:gclspec}
(c) is at $-46 \pm 4$\kms, consistent with that expected if the emission arises from the
3-kpc expanding arm (Section~\ref{sec:gc3kpc}).

The map of Fig.~\ref{fig:gclvelmap} shows the central velocity
distribution of the RRL emission
towards the GC. The morphology of the whole region along with the fact that
the line widths are very similar, suggest that the north and south
lobes are associated, and possibly have the same origin. The velocity gradients, $\sim
15$ -- 18\kms\ from north to south and $\sim2$ -- 5\kms\ from east to
west, could indicate that the structure is in rotation.  
\citet{Law:2009} find that the electron temperature, gas pressure and
morphology of the northern lobe are consistent with it being located
in the GC. Furthermore, they estimate that the three large star clusters,
the Arches, Quintuplet, and Central, in the GC region can account for
the ionization of the northern lobe. These three star clusters are
mainly below the east ridge, whereas the west ridge lies above the Sgr
C \hii~region (G359.4-0.1). \citet{Law:2010}
suggests that the fact that no southern counterpart is observed can be
explained by the offset of the lobes relative to Sgr A$^{\star}$.
Here we detect a southern extension of the GC ionized lobe, which lies
below the source Sgr B, a giant molecular cloud containing
evolved and diffuse \hii~regions as well as young compact and
ultra-compact \hii~regions. This result is in principle consistent
with the outflow origin of the GC lobe, as proposed by \citet{Law:2010}.

\subsection{The ionized 3-kpc arm}
\label{sec:gc3kpc}

 The 3-kpc expanding arm was found in \hi\ lying on the near side of
 the GC by \citet{Rougoor:1964}. At $l = 0\deg$ its velocity was
 $-53$\kms, indicating an outflow.   
A corresponding expanding feature was found in \hi\ by
\citet{Cohen:1976} lying behind the GC with a velocity of
135\kms. Both of these expanding features are seen in CO emission
\citep{Bania:1980}. The velocity of the 3-kpc arm ranges from
$-140$\kms\ at the tangent point at $l = 338\deg$ to $-25$\kms\ at $l
= 10\deg$, where it becomes indistinguishable from the emission of the
normal spiral arms.  

Our RRL survey of the Galactic plane shows only one ``compact''
\hii~region at a velocity expected for the 3-kpc arm (G355.5-0.04). The
scarcity of normal compact \hii~regions in the 3-kpc arm can also be
seen in the H87$\alpha$ (10\,GHz at 3\,arcmin resolution) data of
\citet{Lockman:1989} who found only this same object (G355.7-0.01 at
$V=-76$\kms). Likewise, the southern sky survey \citep{Caswell:1987}
of H109$\alpha$ (5\,GHz at 4\,arcmin resolution) shows only
G359.4-0.09 in the 3-kpc arm. \citet{Cersosimo:1990} found weak
emission in the Galactic centre area in the H166$\alpha$ line at a
resolution of 34\,arcmin. 
\begin{figure*}
\centering
\begin{tabular}{cc}
\includegraphics[scale=0.32]{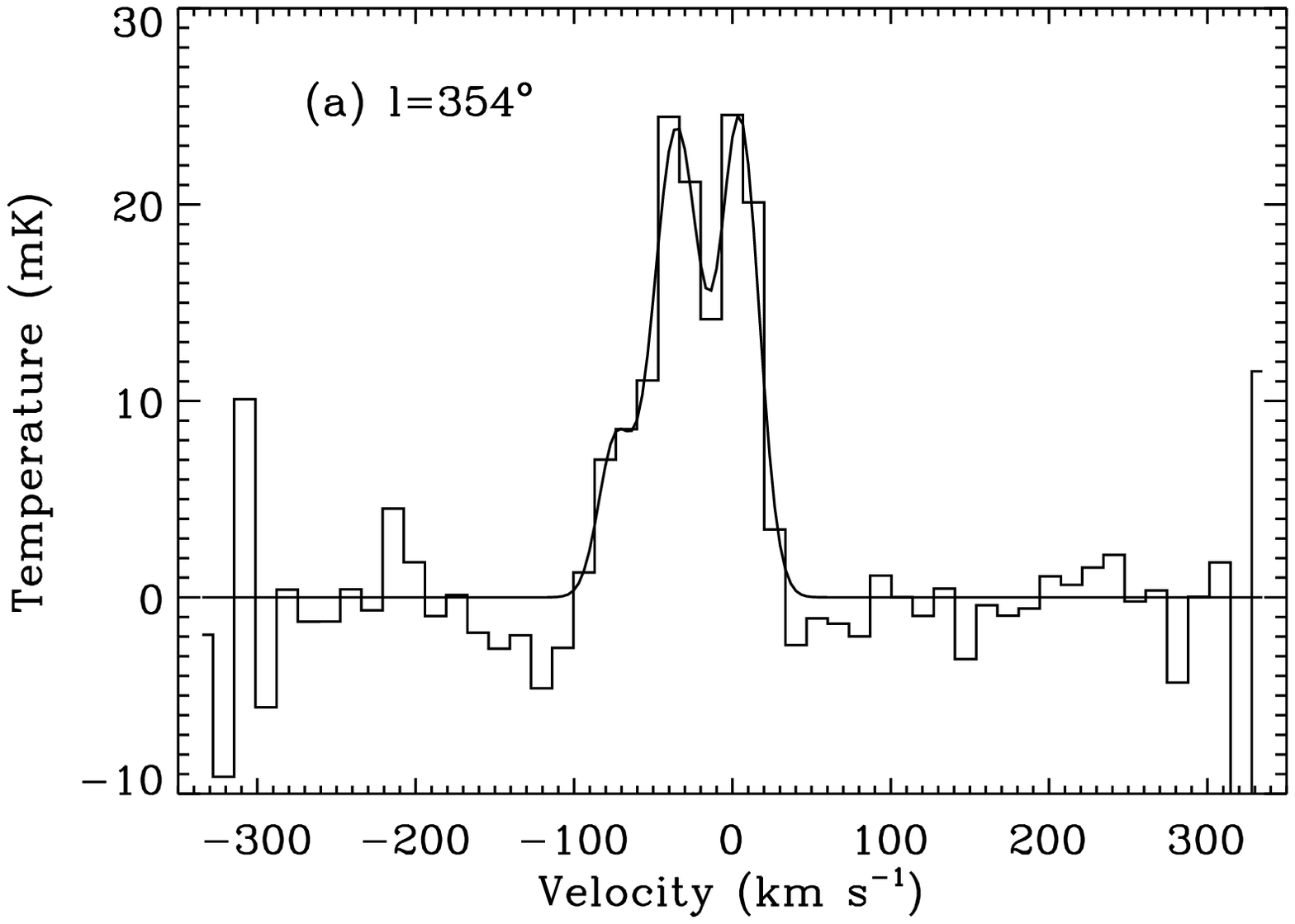}
& \includegraphics[scale=0.32]{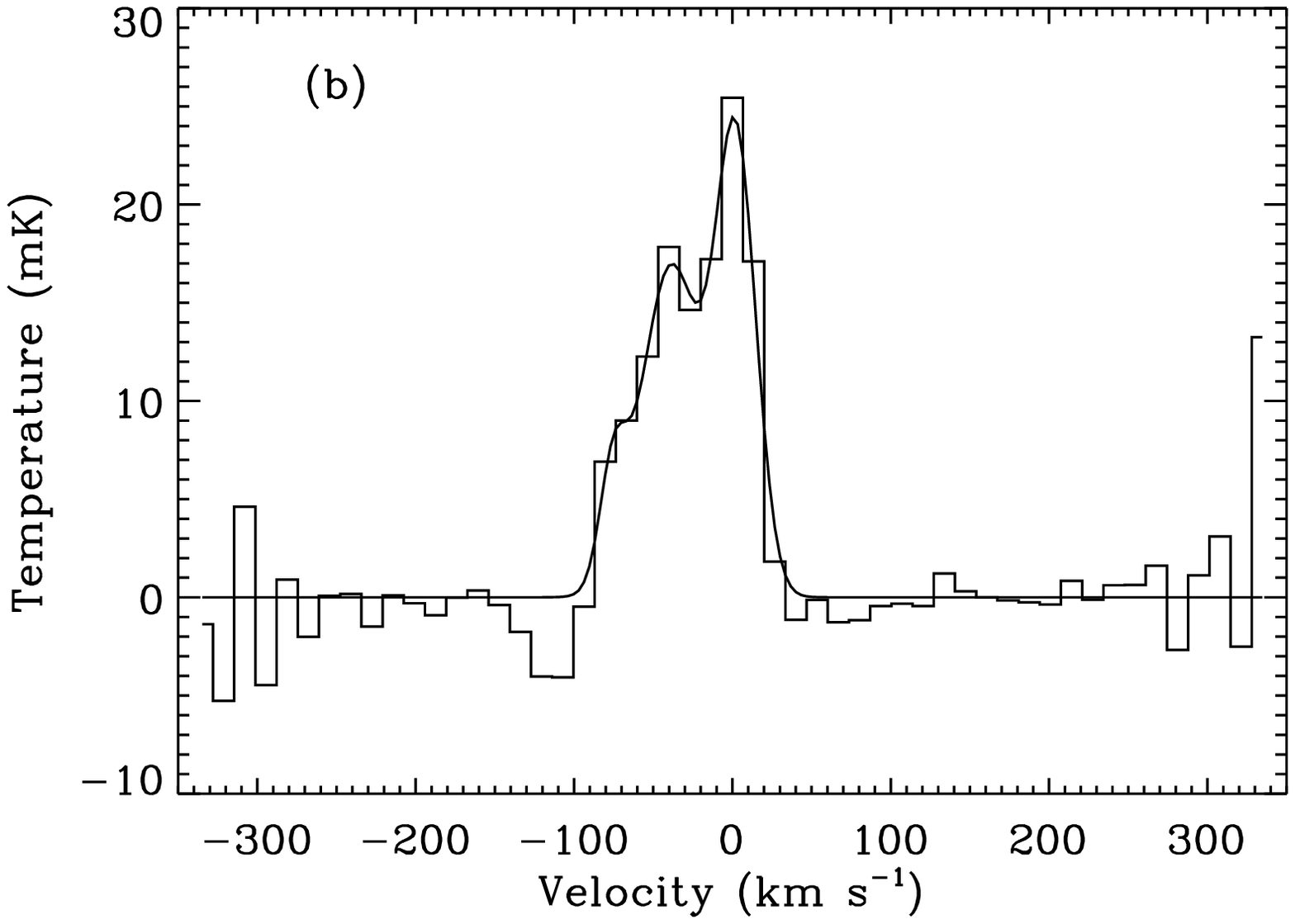} \\
\includegraphics[scale=0.32]{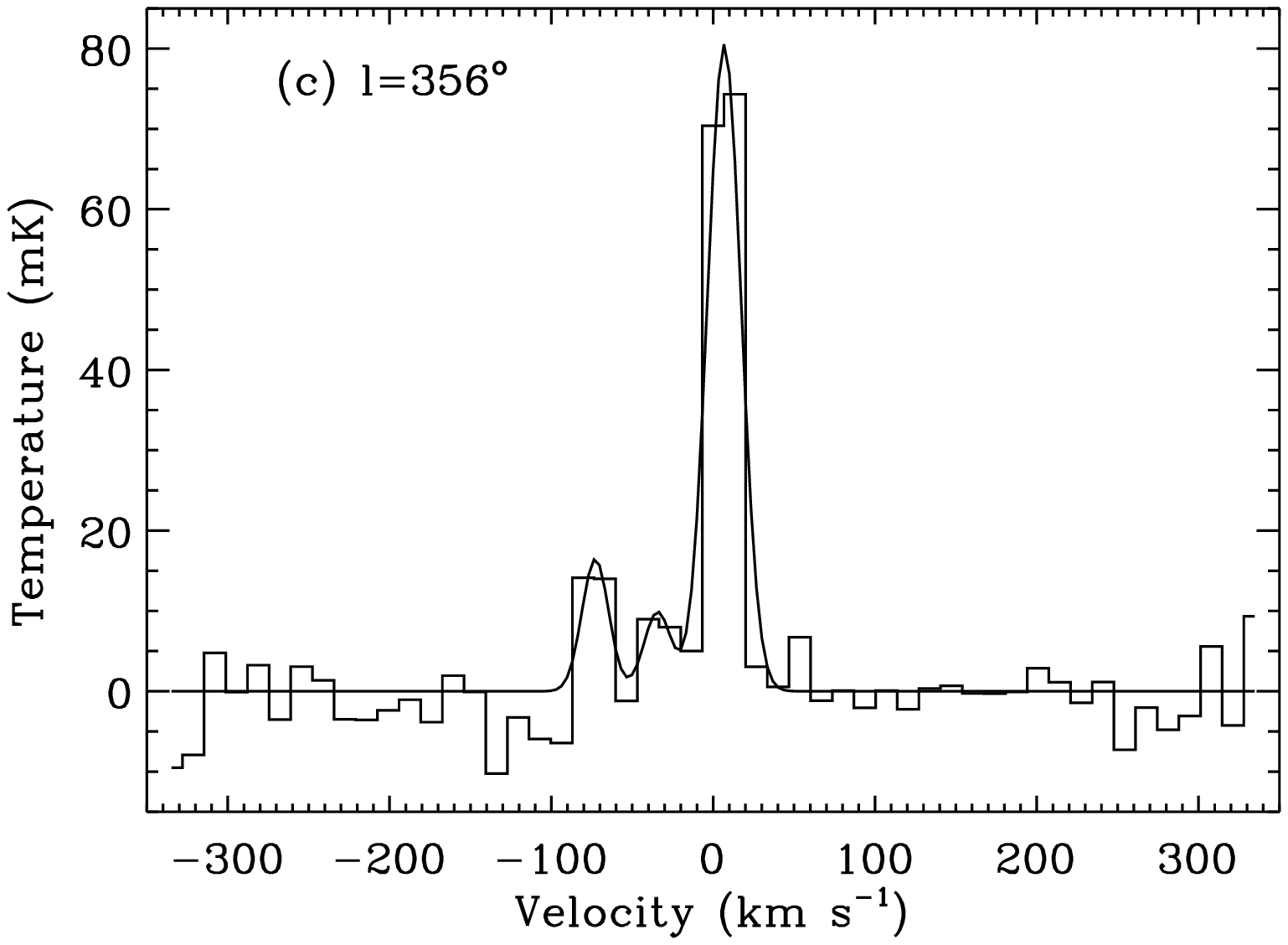} &
\includegraphics[scale=0.32]{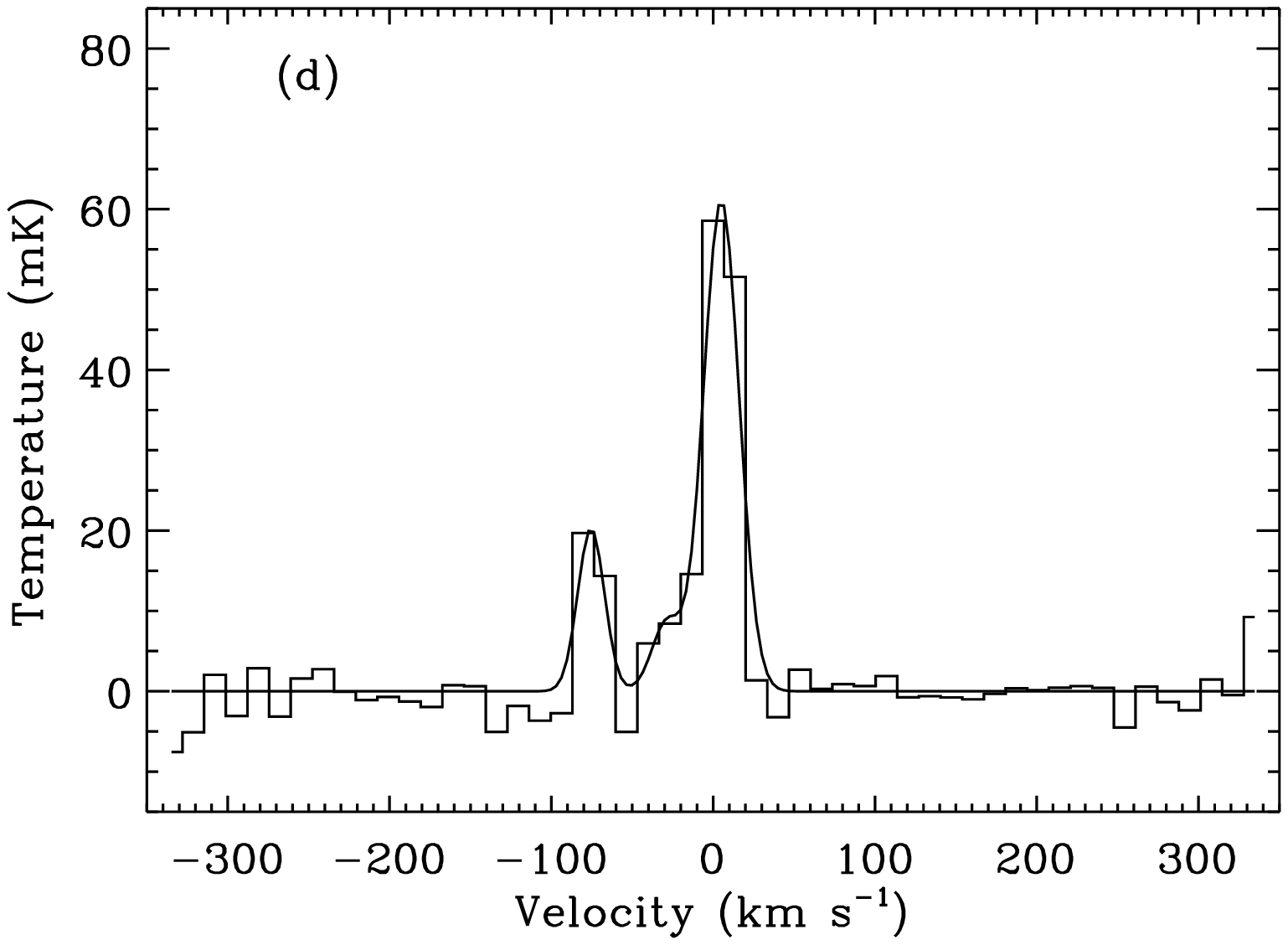} \\
\end{tabular}
\caption{RRL spectra at two positions on the 3-kpc expanding arm, $l =
  354\pdeg0$ and $356\pdeg0$, both at $b = 0\pdeg0$. The spectra on
  the left (panels (a) and (c)) are averaged within a $12 \times 12$\,arcmin$^{2}$ area
  and those on the right (panels (b) and (d)) within $60 \times 60$\,arcmin$^{2}$. Gaussian
  fits of multiple velocity components are shown.\label{fig:3kpcspec}}
\end{figure*}

The spectra near the GC complex shown in Fig. \ref{fig:gclspec}
include the 3-kpc emission as well as the emission in the immediate
vicinity of the centre. A signal of $\sim10$\,mK is seen in some of
the locations. The strongest signal,
45\,mK, is at $(l,b) = (0\pdeg00, 0\pdeg27)$. The diffuse RRL emission
is shown in Fig.~\ref{fig:3kpcspec}, which compares the average
emission in a 12\,arcmin region with that in a 60\,arcmin region
centred at the same position. It can be seen that the average line
temperature in the smaller region is essentially the same as in the
larger. At $(l,b) = (354\pdeg0, 0\pdeg0)$ the expected velocity of the
3-kpc arm is $-80$\kms; it has a temperature of 10\,mK in both 12 and
60 arcmin regions. At $(l,b) = (356\pdeg0, 0\pdeg0)$ the line is
$\sim15$\,mK at the expected velocity of $-71$\kms.

The strength of the present work is the full
coverage of the Galactic plane, which includes not only the individual
\hii~regions but also the diffuse emission. Our work
indicates that $\sim30$ per cent of the total RRL emission
is in individual \hii~regions while 70 per cent is diffuse emission
(Section~\ref{sec:catprop}, \paper). In
the 3-kpc arm the ratio is more like 10:90. This raises the question
of the source of the ionization of this diffuse gas. There is
clearly adequate neutral gas (both \hi\ and CO). The CO is clumped
\citep{Bania:1980} on scales of 20\,arcmin ($\sim50$\,pc) but appears
not to have had enough time to form stars. The origin of the 3-kpc arm remains a mystery.

\section{SUMMARY}
\label{sec:conc}

We have derived the first fully sampled maps of the ionized emission
in the Galactic plane region $l=196\degr$ -- $0\degr$ -- $52\degr$ and $|b| \leq
5\degr$ using the three H$\alpha$ RRLs in the HIPASS/ZOA survey at
1.4\,GHz and 14.4\,arcmin resolution. With a spectral resolution of
20\kms, the rms noise per channel of the stacked line is 4.5\,mK. We
have presented the RRL survey data and discussed their properties and
calibration. 

The RRL maps show the individual \hii~regions as well as diffuse
emission along the inner Galactic plane. We have converted the line
integrated emission to a free-free brightness temperature, which
represents the first direct measure of the thermal emission in
this region of the Galaxy. Following the results of our previous work
(\paper), we have assumed that the electron temperature of the diffuse
ionized gas is similar to that of individual \hii~regions. We find
that the mean electron temperature in the Galactic plane is about
6000\,K. The free-free map is used to extract a catalogue of 317
\hii~regions with flux densities at 1.4\,GHz,
angular sizes and velocities, which can be used to estimate
distances. The individual sources  
account for about 30 per cent of the total free-free emission
in this region of the Galaxy, with 70 per cent being diffuse emission. The ionized gas
arises from the Local, Sagittarius/Carina and Scutum/Norma spiral
arms, but is mostly located in the molecular
ring. Within these inner 30\degr\ of longitude, there is a wider
velocity spread in the CO line than in the RRLs, illustrating the
localised star formation in the Galaxy. We have also presented RRL
observations towards the Galactic centre, where a previously
identified lobe of ionized gas north of the Galactic plane was
associated with a mass outflow. These observations provide the first
detection, in RRLs, of its southern counterpart and the indication that
both degree-scale structures are related. The present data also
provide further evidence of diffuse ionized gas in the expanding 3-kpc
arm, which had only previously been reported for a few individual lines of
sight. We have also studied the distribution of helium and carbon RRLs in the
Scutum spiral arm, finding diffuse C{\sc ii} gas emission
around \hii~complexes in the Galactic plane. 


The following RRL HIPASS/ZOA data products are available at
\textcolor{red}{\url{http://www.jodrellbank.manchester.ac.uk/research/parkes_rrl_survey/}}:
\begin{itemize}
\item The RRL datacube covering the whole spatial extent of the
  survey, $(\Delta l, \Delta b)=(216\degr, 10\degr)$, with a $V_{\rm
    LSR}$ coverage of $\pm 335$\kms, spectral resolution of 20\kms,
  channel width of 13.4\kms. This corresponds to the combination of
  the H$168\alpha$, H$167\alpha$, and H$166\alpha$ RRLs and is in
  units of mK brightness temperature, with a typical rms noise per channel of
  4.5\,mK. 
\item The RRL integrated map in units of K\,km\,s$^{-1}$. 
\item The electron temperature map in units of K. 
\item The free-free emission map in units of K brightness temperature
  at 1.4\,GHz, estimated with an electron temperature gradient with Galactocentric radius of
$496\pm100$\,K\,kpc$^{-1}$. 
\end{itemize}
All the data are at an angular resolution of 14.4\,arcmin and calibrated on the full beam scale. 

\section*{ACKNOWLEDGEMENTS}

We thank the referee, F. J. Lockman, for the helpful comments.
MIRA acknowledges the support by the European Research Council grant
MISTIC (ERC-267934). CD acknowledges support from an STFC Advanced
Fellowship, an EU Marie-Cure IRG grant under the FP7, and an ERC
Starting (Consolidator) Grant (no.~307209). The Parkes telescope is part of the
Australia Telescope which is funded by the Commonwealth of Australia
for operation as a National Facility managed by CSIRO. The Southern
H-Alpha Sky Survey Atlas (SHASSA) is supported by the National Science
Foundation.


\bibliographystyle{mn2e}
\bibliography{refs1}

\begin{thebibliography}{71}
\expandafter\ifx\csname natexlab\endcsname\relax\def\natexlab#1{#1}\fi

\bibitem[{{Alves} {et~al}\mbox{.}(2012){Alves}, {Davies}, {Dickinson},
  {Calabretta}, {Davis}, \& {Staveley-Smith}}]{Alves:2012}
{Alves} M.~I.~R., {Davies} R.~D., {Dickinson} C., {Calabretta} M., {Davis} R.,
  {Staveley-Smith} L., 2012, \mnras, 422, 2429

\bibitem[{{Alves} {et~al}\mbox{.}(2010){Alves}, {Davies}, {Dickinson}, {Davis},
  {Auld}, {Calabretta}, \& {Staveley-Smith}}]{Alves:2010}
{Alves} M.~I.~R., {Davies} R.~D., {Dickinson} C., {Davis} R.~J., {Auld} R.~R.,
  {Calabretta} M., {Staveley-Smith} L., 2010, \mnras, 405, 1654

\bibitem[{{Anderson} {et~al}\mbox{.}(2011){Anderson}, {Bania}, {Balser}, \&
  {Rood}}]{Anderson:2011}
{Anderson} L.~D., {Bania} T.~M., {Balser} D.~S., {Rood} R.~T., 2011, \apjs,
  194, 32

\bibitem[{{Baddi}(2012)}]{Baddi:2012}
{Baddi} R., 2012, \aj, 143, 45

\bibitem[{{Balick}, {Gammon} \& {Hjellming}(1974){Balick}, {Gammon}, \&
  {Hjellming}}]{Balick:1974}
{Balick} B., {Gammon} R.~H., {Hjellming} R.~M., 1974, \pasp, 86, 616

\bibitem[{{Bania}(1980)}]{Bania:1980}
{Bania} T.~M., 1980, \apj, 242, 95

\bibitem[{{Barnes} {et~al}\mbox{.}(2001){Barnes}, {Staveley-Smith}, {de Blok},
  {Oosterloo}, {Stewart}, {Wright}, {Banks}, {Bhathal}, {Boyce}, {Calabretta},
  {Disney}, {Drinkwater}, {Ekers}, {Freeman}, {Gibson}, {Green}, {Haynes}, {te
  Lintel Hekkert}, {Henning}, {Jerjen}, {Juraszek}, {Kesteven}, {Kilborn},
  {Knezek}, {Koribalski}, {Kraan-Korteweg}, {Malin}, {Marquarding}, {Minchin},
  {Mould}, {Price}, {Putman}, {Ryder}, {Sadler}, {Schr{\"o}der}, {Stootman},
  {Webster}, {Wilson}, \& {Ye}}]{Barnes:2001}
{Barnes} D.~G. {et~al.}, 2001, \mnras, 322, 486

\bibitem[{{Bertin} \& {Arnouts}(1996)}]{Bertin:1996}
{Bertin} E., {Arnouts} S., 1996, \aaps, 117, 393

\bibitem[{{Bihr} {et~al}\mbox{.}(2013){Bihr}, {Beuther}, {Johnston}, {Ott},
  {Glover}, {Carlhoff}, {Brunthaler}, {Goldsmith}, {Schilke}, {Motte}, \&
  {Henning}}]{THOR:2013}
{Bihr} S. {et~al.}, 2013, in Protostars and Planets VI Posters, p.~50

\bibitem[{{Bronfman} {et~al}\mbox{.}(2000){Bronfman}, {Casassus}, {May}, \&
  {Nyman}}]{Bronfman:2000}
{Bronfman} L., {Casassus} S., {May} J., {Nyman} L.-{\AA}., 2000, \aap, 358, 521

\bibitem[{{Burton}(1988)}]{Burton:1988}
{Burton} W.~B., 1988, in K.~Kellerman, G. L.~Verschuur, ed., Galactic and
  Extragalactic Radio Astronomy. New York: Springer-Verlag

\bibitem[{{Burton} \& {Hekkert}(1985)}]{Burton:1985a}
{Burton} W.~B., {Hekkert} P.~T.~L., 1985, \aaps, 62, 645

\bibitem[{{Calabretta}, {Staveley-Smith} \& {Barnes}(2014){Calabretta},
  {Staveley-Smith}, \& {Barnes}}]{Calabretta:2014}
{Calabretta} M.~R., {Staveley-Smith} L., {Barnes} D.~G., 2014, PASA, 31, 7

\bibitem[{{Caswell} \& {Haynes}(1987)}]{Caswell:1987}
{Caswell} J.~L., {Haynes} R.~F., 1987, \aap, 171, 261

\bibitem[{{Cersosimo} \& {Magnani}(1990)}]{Cersosimo:1990}
{Cersosimo} J.~C., {Magnani} L., 1990, \aap, 239, 287

\bibitem[{{Cohen} \& {Davies}(1976)}]{Cohen:1976}
{Cohen} R.~J., {Davies} R.~D., 1976, \mnras, 175, 1

\bibitem[{{Dame}, {Hartmann} \& {Thaddeus}(2001){Dame}, {Hartmann}, \&
  {Thaddeus}}]{Dame:2001}
{Dame} T.~M., {Hartmann} D., {Thaddeus} P., 2001, \apj, 547, 792

\bibitem[{{Dame} \& {Thaddeus}(2008)}]{Dame:2008}
{Dame} T.~M., {Thaddeus} P., 2008, \apjl, 683, L143

\bibitem[{{Dickinson}, {Davies} \& {Davis}(2003){Dickinson}, {Davies}, \&
  {Davis}}]{D3:2003}
{Dickinson} C., {Davies} R.~D., {Davis} R.~J., 2003, \mnras, 341, 369

\bibitem[{{Draine}(2011)}]{Draine:2011}
{Draine} B.~T., 2011, {Physics of the Interstellar and Intergalactic Medium
  (Princeton University Press)}

\bibitem[{{Dupree}(1974)}]{Dupree:1974}
{Dupree} A.~K., 1974, \apj, 187, 25

\bibitem[{{Erickson}, {McConnell} \& {Anantharamaiah}(1995){Erickson},
  {McConnell}, \& {Anantharamaiah}}]{Erickson:1995}
{Erickson} W.~C., {McConnell} D., {Anantharamaiah} K.~R., 1995, \apj, 454, 125

\bibitem[{{Evans}(1991)}]{Evans:1991}
{Evans}, II N.~J., 1991, in Astronomical Society of the Pacific Conference
  Series, Vol.~20, Frontiers of Stellar Evolution, {Lambert} D.~L., ed., pp.
  45--95

\bibitem[{{Fich}, {Blitz} \& {Stark}(1989){Fich}, {Blitz}, \&
  {Stark}}]{Fich:1989}
{Fich} M., {Blitz} L., {Stark} A.~A., 1989, \apj, 342, 272

\bibitem[{{Finkbeiner}(2003)}]{Finkbeiner:2003}
{Finkbeiner} D.~P., 2003, \apjs, 146, 407

\bibitem[{{Gaustad} {et~al}\mbox{.}(2001){Gaustad}, {McCullough}, {Rosing}, \&
  {Van Buren}}]{Gaustad:2001}
{Gaustad} J.~E., {McCullough} P.~R., {Rosing} W., {Van Buren} D., 2001, \pasp,
  113, 1326

\bibitem[{{Goldberg} \& {Dupree}(1967)}]{Goldberg:1967}
{Goldberg} L., {Dupree} A.~K., 1967, \nat, 215, 41

\bibitem[{{Gottesman} \& {Gordon}(1970)}]{Gottesman:1970}
{Gottesman} S.~T., {Gordon} M.~A., 1970, \apjl, 162, L93

\bibitem[{{Hart} \& {Pedlar}(1976)}]{Hart:1976}
{Hart} L., {Pedlar} A., 1976, \mnras, 176, 547

\bibitem[{{Haslam} {et~al}\mbox{.}(1982){Haslam}, {Salter}, {Stoffel}, \&
  {Wilson}}]{Haslam:1982}
{Haslam} C.~G.~T., {Salter} C.~J., {Stoffel} H., {Wilson} W.~E., 1982, \aaps,
  47, 1

\bibitem[{{Haynes}, {Caswell} \& {Simons}(1978){Haynes}, {Caswell}, \&
  {Simons}}]{Haynes:1978}
{Haynes} R.~F., {Caswell} J.~L., {Simons} L.~W.~J., 1978, Australian Journal of
  Physics Astrophysical Supplement, 45, 1

\bibitem[{{Heiles}, {Reach} \& {Koo}(1996){Heiles}, {Reach}, \&
  {Koo}}]{Heiles:1996}
{Heiles} C., {Reach} W.~T., {Koo} B., 1996, \apj, 466, 191

\bibitem[{{Jackson} \& {Kerr}(1971)}]{Jackson:1971}
{Jackson} P.~D., {Kerr} F.~J., 1971, \apj, 168, 29

\bibitem[{{Jarosik} {et~al}\mbox{.}(2011){Jarosik}, {Bennett}, {Dunkley},
  {Gold}, {Greason}, {Halpern}, {Hill}, {Hinshaw}, {Kogut}, {Komatsu},
  {Larson}, {Limon}, {Meyer}, {Nolta}, {Odegard}, {Page}, {Smith}, {Spergel},
  {Tucker}, {Weiland}, {Wollack}, \& {Wright}}]{Jarosik:2011}
{Jarosik} N. {et~al.}, 2011, \apjs, 192, 14

\bibitem[{{Jonas}, {Baart} \& {Nicolson}(1998){Jonas}, {Baart}, \&
  {Nicolson}}]{Jonas:1998}
{Jonas} J.~L., {Baart} E.~E., {Nicolson} G.~D., 1998, \mnras, 297, 977

\bibitem[{{Kantharia} \& {Anantharamaiah}(2001)}]{Kantharia:2001}
{Kantharia} N.~G., {Anantharamaiah} K.~R., 2001, Journal of Astrophysics and
  Astronomy, 22, 51

\bibitem[{{Kerr} \& {Lynden-Bell}(1986)}]{Kerr:1986}
{Kerr} F.~J., {Lynden-Bell} D., 1986, \mnras, 221, 1023

\bibitem[{{Kuchar} \& {Clark}(1997)}]{Kuchar:1997}
{Kuchar} T.~A., {Clark} F.~O., 1997, \apj, 488, 224

\bibitem[{{Law}(2010)}]{Law:2010}
{Law} C.~J., 2010, \apj, 708, 474

\bibitem[{{Law} {et~al}\mbox{.}(2009){Law}, {Backer}, {Yusef-Zadeh}, \&
  {Maddalena}}]{Law:2009}
{Law} C.~J., {Backer} D., {Yusef-Zadeh} F., {Maddalena} R., 2009, \apj, 695,
  1070

\bibitem[{{Liu} {et~al}\mbox{.}(2013){Liu}, {McIntyre}, {Terzian}, {Minchin},
  {Anderson}, {Churchwell}, {Lebron}, \& {Anish Roshi}}]{Liu:2013}
{Liu} B., {McIntyre} T., {Terzian} Y., {Minchin} R., {Anderson} L.,
  {Churchwell} E., {Lebron} M., {Anish Roshi} D., 2013, \aj, 146, 80

\bibitem[{{Lockman}(1976)}]{Lockman:1976}
{Lockman} F.~J., 1976, \apj, 209, 429

\bibitem[{{Lockman}(1989)}]{Lockman:1989}
---, 1989, \apjs, 71, 469

\bibitem[{{Mezger} {et~al}\mbox{.}(1967){Mezger}, {Altenhoff}, {Schraml},
  {Burke}, {Reifenstein}, \& {Wilson}}]{Mezger:1967}
{Mezger} P.~G., {Altenhoff} W., {Schraml} J., {Burke} B.~F., {Reifenstein}, III
  E.~C., {Wilson} T.~L., 1967, \apjl, 150, L157

\bibitem[{{Miville-Desch{\^e}nes} \& {Lagache}(2005)}]{Miville-Deschenes:2005}
{Miville-Desch{\^e}nes} M., {Lagache} G., 2005, \apjs, 157, 302

\bibitem[{{Paladini} {et~al}\mbox{.}(2003){Paladini}, {Burigana}, {Davies},
  {Maino}, {Bersanelli}, {Cappellini}, {Platania}, \& {Smoot}}]{Paladini:2003}
{Paladini} R., {Burigana} C., {Davies} R.~D., {Maino} D., {Bersanelli} M.,
  {Cappellini} B., {Platania} P., {Smoot} G., 2003, \aap, 397, 213

\bibitem[{{Paladini}, {Davies} \& {DeZotti}(2004){Paladini}, {Davies}, \&
  {DeZotti}}]{Paladini:2004}
{Paladini} R., {Davies} R.~D., {DeZotti} G., 2004, \mnras, 347, 237

\bibitem[{{Paladini} {et~al}\mbox{.}(2005){Paladini}, {De Zotti}, {Davies}, \&
  {Giard}}]{Paladini:2005}
{Paladini} R., {De Zotti} G., {Davies} R.~D., {Giard} M., 2005, \mnras, 360,
  1545

\bibitem[{{Planck Collaboration}(2014{\natexlab{a}})}]{PIP96}
{Planck Collaboration}, 2014{\natexlab{a}}, \aap, 564, A45

\bibitem[{{Planck Collaboration}(2014{\natexlab{b}})}]{PIP79}
---, 2014{\natexlab{b}}, submitted to A\&A, [arXiv:1406.5093]

\bibitem[{{Reich} \& {Reich}(1986)}]{Reich:1986}
{Reich} P., {Reich} W., 1986, \aaps, 63, 205

\bibitem[{{Reich}, {Testori} \& {Reich}(2001){Reich}, {Testori}, \&
  {Reich}}]{Reich:2001}
{Reich} P., {Testori} J.~C., {Reich} W., 2001, \aap, 376, 861

\bibitem[{{Reich}(1982)}]{Reich:1982}
{Reich} W., 1982, \aaps, 48, 219

\bibitem[{{Reich} {et~al}\mbox{.}(1990){Reich}, {F\"{u}erst}, {Reich}, \&
  {Reif}}]{Reich:1990a}
{Reich} W., {F\"{u}erst} E., {Reich} P., {Reif} K., 1990, \aaps, 85, 633

\bibitem[{{Rohlfs} \& {Wilson}(2000)}]{Rohlfs:2000}
{Rohlfs} K., {Wilson} T.~L., 2000, in K.~Rohlfs, T.L.~Wilson, ed., Tools of
  Radio Astronomy. New York: Springer

\bibitem[{{Roshi} \& {Anantharamaiah}(2000)}]{Roshi:2000}
{Roshi} D.~A., {Anantharamaiah} K.~R., 2000, \apj, 535, 231

\bibitem[{{Roshi}, {Kantharia} \& {Anantharamaiah}(2002){Roshi}, {Kantharia},
  \& {Anantharamaiah}}]{Roshi:2002}
{Roshi} D.~A., {Kantharia} N.~G., {Anantharamaiah} K.~R., 2002, \aap, 391, 1097

\bibitem[{{Rougoor}(1964)}]{Rougoor:1964}
{Rougoor} G.~W., 1964, BAN, 17, 381

\bibitem[{{Sault}, {Teuben} \& {Wright}(1995){Sault}, {Teuben}, \&
  {Wright}}]{Sault:1995}
{Sault} R.~J., {Teuben} P.~J., {Wright} M.~C.~H., 1995, in Astronomical Society
  of the Pacific Conference Series, Vol.~77, Astronomical Data Analysis
  Software and Systems IV, {Shaw} R.~A., {Payne} H.~E., {Hayes} J.~J.~E., eds.,
  p. 433

\bibitem[{{Shaver} {et~al}\mbox{.}(1983){Shaver}, {McGee}, {Newton}, {Danks},
  \& {Pottasch}}]{Shaver:1983}
{Shaver} P.~A., {McGee} R.~X., {Newton} L.~M., {Danks} A.~C., {Pottasch} S.~R.,
  1983, \mnras, 204, 53

\bibitem[{{Smith} \& {Brooks}(2008)}]{Smith:2008}
{Smith} N., {Brooks} K.~J., 2008, {The Carina Nebula: A Laboratory for Feedback
  and Triggered Star Formation}, {Reipurth} B., ed., p. 138

\bibitem[{{Sofue} \& {Handa}(1984)}]{Sofue:1984}
{Sofue} Y., {Handa} T., 1984, \nat, 310, 568

\bibitem[{{Staveley-Smith} {et~al}\mbox{.}(1998){Staveley-Smith}, {Juraszek},
  {Koribalski}, {Ekers}, {Green}, {Haynes}, {Henning}, {Kesteven},
  {Kraan-Korteweg}, {Price}, \& {Sadler}}]{Staveley-Smith:1998}
{Staveley-Smith} L. {et~al.}, 1998, \aj, 116, 2717

\bibitem[{{Staveley-Smith} {et~al}\mbox{.}(2003){Staveley-Smith}, {Kim},
  {Calabretta}, {Haynes}, \& {Kesteven}}]{Staveley-Smith:2003}
{Staveley-Smith} L., {Kim} S., {Calabretta} M.~R., {Haynes} R.~F., {Kesteven}
  M.~J., 2003, \mnras, 339, 87

\bibitem[{{Staveley-Smith} {et~al}\mbox{.}(1996){Staveley-Smith}, {Wilson},
  {Bird}, {Disney}, {Ekers}, {Freeman}, {Haynes}, {Sinclair}, {Vaile},
  {Webster}, \& {Wright}}]{Staveley-Smith:1996}
{Staveley-Smith} L. {et~al.}, 1996, Publications of the Astronomical Society of
  Australia, 13, 243

\bibitem[{{Stil} {et~al}\mbox{.}(2006){Stil}, {Taylor}, {Dickey}, {Kavars},
  {Martin}, {Rothwell}, {Boothroyd}, {Lockman}, \&
  {McClure-Griffiths}}]{Stil:2006}
{Stil} J.~M. {et~al.}, 2006, \aj, 132, 1158

\bibitem[{{Sun} {et~al}\mbox{.}(2011){Sun}, {Reich}, {Han}, {Reich},
  {Wielebinski}, {Wang}, \& {M{\"u}ller}}]{Sun:2011}
{Sun} X.~H., {Reich} W., {Han} J.~L., {Reich} P., {Wielebinski} R., {Wang} C.,
  {M{\"u}ller} P., 2011, \aap, 527, A74

\bibitem[{{Tibbs} {et~al}\mbox{.}(2012){Tibbs}, {Paladini}, {Compi{\`e}gne},
  {Dickinson}, {Alves}, {Flagey}, {Shenoy}, {Noriega-Crespo}, {Carey},
  {Casassus}, {Davies}, {Davis}, {Molinari}, {Elia}, {Pestalozzi}, \&
  {Schisano}}]{Tibbs:2012}
{Tibbs} C.~T. {et~al.}, 2012, \apj, 754, 94

\bibitem[{{Traficante} {et~al}\mbox{.}(2014){Traficante}, {Paladini},
  {Compiegne}, {Alves}, {Cambr{\'e}sy}, {Gibson}, {Tibbs}, {Noriega-Crespo},
  {Molinari}, {Carey}, {Ingalls}, {Natoli}, {Davies}, {Davis}, {Dickinson}, \&
  {Fuller}}]{Traficante:2014}
{Traficante} A. {et~al.}, 2014, \mnras, 440, 3588

\bibitem[{Tukey(1967)}]{Tukey:1967}
Tukey J.~W., 1967, in B. Harris, ed., Spectral Analysis of Time Series. New
  York: Wiley, 25

\bibitem[{{Wood} \& {Churchwell}(1989)}]{Wood:1989}
{Wood} D.~O.~S., {Churchwell} E., 1989, \apj, 340, 265

\end{thebibliography}

\raggedright


\bsp 

\label{lastpage}

\end{document}